\documentclass[aps,preprint,nofootinbib,preprintnumbers,eqsecnum]{revtex4-1}

\usepackage{color}

\usepackage[
      colorlinks=true,
      linkcolor=blue,
      urlcolor=blue,
      filecolor=black,
      citecolor=red,
      pdfstartview=FitV,
      pdftitle={},
        pdfauthor={Thomas Faulkner, Hong Liu, Mukund Rangamani},
        pdfsubject={},
        pdfkeywords={},
        pdfpagemode=None,
        bookmarksopen=true
      ]{hyperref}

\usepackage[normalem]{ulem}
\usepackage{amsmath}
\usepackage{enumerate}
\usepackage{amsfonts}
\usepackage{yfonts}

\usepackage{subfigure}
\usepackage{psfrag}

\usepackage{epsfig}
\usepackage[latin1]{inputenc}
\usepackage{float}
\usepackage{graphicx}
\usepackage{cancel}
\usepackage{mathrsfs}
\usepackage{amssymb}
\usepackage{amsfonts}
\usepackage{amsmath}
\usepackage{slashed}

\newcommand \tr {\mbox{{\bf Tr}}}

\usepackage{graphicx}
\usepackage{bm}

\def\({\left(}
\def\){\right)}
\def\[{\left[}
\def\]{\right]}
\def\<{\langle}
\def\>{\rangle}

\def\Tr{\mathop{\rm Tr}}
\def\tr{\mathop{\rm tr}}

\newcommand\half{{\ensuremath{\frac{1}{2}}}}
\newcommand\p{\ensuremath{\partial}}

\newcommand\field[1]{{\ensuremath{\mathbb{{#1}}}}}

\newcommand\vev[1]{{\ensuremath{\left\langle{#1}\right\rangle}}}

\newcommand{\RR}{\field{R}}

\newcommand{\be}{\begin{equation}}
\newcommand{\ee}{\end{equation}}
\newcommand{\bea}{\begin{eqnarray}}
\newcommand{\eea}{\end{eqnarray}}
\newcommand{\bwt}{\begin{widetext}}
\newcommand{\ewt}{\end{widetext}}

\newcommand{\bi}{\begin{itemize}}
\newcommand{\ei}{\end{itemize}}
\newcommand{\ben}{\begin{enumerate}}
\newcommand{\een}{\end{enumerate}}
\newcommand{\bca}{\begin{cases}}
\newcommand{\eca}{\end{cases}}
\newcommand{\bln}{\begin{align}}
\newcommand{\eln}{\end{align}}
\newcommand{\bst}{\begin{split}}
\newcommand{\est}{\end{split}}

\newcommand\al{{\alpha}}
\newcommand\ep{\epsilon}
\newcommand\sig{\sigma}
\newcommand\Sig{\Sigma}

\newcommand\Lam{\Lambda}
\newcommand\om{\omega}
\newcommand\Om{\Omega}

\newcommand\ga{{\ensuremath{{\gamma}}}}
\newcommand\Ga{{\ensuremath{{\Gamma}}}}
\newcommand\de{{\ensuremath{{\delta}}}}
\newcommand\De{{\ensuremath{{\Delta}}}}

\def\th{{\theta}}

\newcommand\ov{\over}
\newcommand\ha{{\half}}

\def\le{\left}
\def\ri{\right}

\newcommand\sH{{\ensuremath{{\mathcal H}}}}
\newcommand\sL{{\ensuremath{{\mathcal L}}}}

\newcommand\sN{{\ensuremath{{\mathcal N}}}}
\newcommand\sO{{\ensuremath{{\mathcal O}}}}
\newcommand\sR{{\ensuremath{{\mathcal R}}}}

\newcommand\sS{{\mathcal S}}

\newcommand{\ka}{{\kappa}}

\newcommand{\sSS}{\sS^{(\Sig)}_d (R)}

\newcommand{\sds}{s^{(\Sig)}_d}

\begin{document}

\title {A refinement of entanglement entropy and the number of degrees of freedom
}

\preprint{MIT-CTP 4336}

\author{ Hong Liu}
\affiliation{Center for Theoretical Physics,
Massachusetts
Institute of Technology,
Cambridge, MA 02139 }
\author{M\'ark Mezei}
\affiliation{Center for Theoretical Physics, Massachusetts Institute of Technology,
Cambridge, MA 02139 }

\begin{abstract}

We introduce a ``renormalized entanglement entropy''  which is intrinsically UV finite 
and is most sensitive  to the degrees of freedom at the scale of the size $R$ of the entangled region. 
We illustrated the power of this construction by showing that  the qualitative behavior of the entanglement entropy for a non-Fermi liquid can be obtained by simple dimensional analysis. 
We argue that the functional dependence of the ``renormalized entanglement entropy'' on $R$ can be interpreted as describing the renormalization group flow of the entanglement entropy with distance scale. 
The corresponding quantity for a spherical region in the vacuum, 
has some particularly interesting properties. 
For a conformal field theory, it  reduces to 
the previously proposed central charge in all dimensions, and for a general quantum field theory,  it interpolates between the central charges of the UV and IR fixed points as $R$ is varied from zero to infinity.  We conjecture that in three (spacetime) dimensions, it is always non-negative and monotonic,  
and provides a measure of the number of degrees of freedom of a system at scale $R$.
In four dimensions, however, we find examples in which it is neither monotonic nor non-negative.

\end{abstract}

\today

\maketitle

\tableofcontents


\section{Introduction}

Quantum entanglement has been seen to play an increasingly important role in our understanding and characterization of many-body physics (see e.g.~\cite{amico,eisert}). 
The entanglement entropy for a spatial region provides 
an important set of observables to probe such quantum correlations. 

In spacetime dimensions higher than two, however, the entanglement entropy for a spatial region is dominated by contributions from non-universal, cutoff-scale physics~\cite{Bombelli:1986rw,Srednicki:1993im}.  This implies that for a region characterized by a size $R$, the entanglement entropy is sensitive to the physics from scale $R$ all the way down to the cutoff scale $\de$, no matter how large $R$ is. 
As a result the entanglement entropy is ill-defined in the continuum limit. 
The common practice is to subtract the UV divergent part by hand, a procedure which is not unique and often ambiguous, in particular in systems with more than one scales. 
Even with the UV divergent part removed, the resulting expression could still depend sensitively on physics at scales much smaller than the size $R$ of the entangled region. As a result, in the limit of taking $R$ to infinity, one often does not recover the expected behavior of the IR fixed point~(see for example the case of a free massive scalar in Sec.~\ref{sec:free}). 

Such a situation is clearly awkward both operationally and conceptually. We should be able to probe and characterize quantum entanglement at a given macroscopic scale without worrying about 
physics at much shorter distance scales.

In this paper we show that there is a simple fix of the problem.\footnote{See also~\cite{Hertzberg:2010uv} for a discussion based on free theories.} 
Consider a quantum field theory on $\RR^{1,d-1}$ which is renormalizable non-perturbatively, i.e., equipped with a well-defined UV fixed point.  
Suppose $S^{(\Sig)} (R)$ is the entanglement entropy in the vacuum across some smooth entangling surface $\Sig$ characterized by a scalable size $R$.\footnote{In this paper we will always consider $\Sig$ to be a closed connected surface.  
Also note that not all closed surfaces have a scalable size. In Sec.~\ref{app:UV} and Appendix~\ref{app:ent} we make this more precise.} 
We introduce the following function 
 \vskip-0.7cm
 \be \label{Scen}
 \sS_d^{(\Sig)} (R) = \bca
  {1 \ov  (d-2)!!}  \le(R {d \ov dR} - 1 \ri) \le(R {d \ov dR} -3 \ri)  \cdots \le(R {d \ov dR} -(d-2) \ri) S^{(\Sig)}(R) & {\rm d \; odd} \cr
  {1 \ov  (d-2)!!}  R {d \ov dR}  \le(R {d \ov dR} -2 \ri)  \cdots \le(R {d \ov dR} -(d-2) \ri) S^{(\Sig)}(R) & {\rm d \; even}
    \eca \ .
    \ee
In Sec.~\ref{app:UV} we show that it has the following properties:   
\ben 

\item  It is UV finite in the continuum limit (i.e. when the short-distance cutoff is taken to zero).  

\item For a CFT it is given by a $R$-independent constant  $\sds$. \label{im:2}

\item For a renormalizable quantum field theory, it 
interpolates between the values $s^{(\rm \Sig,UV)}_d$ and $s^{(\rm \Sig,IR)}_d$ of the UV and IR fixed points as $R$ is increased from zero to infinity.  
\label{im:3}

\item It is most sensitive to degrees of freedom at scale $R$. \label{im:4}

\een 

The differential operator in~\eqref{Scen} plays the role of stripping from $S^{(\Sig)} (R)$ of short-distance correlations. The stripping includes also finite subtractions and is $R$-dependent; it gets of rid of not only the UV divergences, but also contributions from degrees of freedom at scales much smaller than $R$. $\sSS$ can be also be used at a finite temperature or  finite density where it is again UV finite in the continuum limit.  In the small $R$ limit it reduces to the vacuum behavior while for large $R$ we expect it to go over to the thermal entropy.

$\sSS$ may be considered as the ``universal part'' of the original entanglement entropy, a part which can be defined intrinsically in the continuum limit. Below we will sometimes refer to it as the ``renormalized entanglement entropy,'' although this name is clearly not perfect. We believe such a 
construction gives a powerful tool for understanding entanglement of a many-body system. 
As an illustration, in Sec.~\ref{sec:fermi} we show that the entanglement entropy of a non-Fermi liquid also has a logarithmic enhancement just as that for a Fermi liquid  by a simple dimensional analysis.
We also predict the behavior of the entanglement entropy from higher co-dimensional Fermi surfaces. 

In the rest of the paper we focus on the behavior of $\sS_d^{(\Sig)} (R)$ in the vacuum, studying its possible connections to renormalization group flow (RG) and the number of degrees of freedom along the flow.  Items~\eqref{im:2}--\eqref{im:4} above, especially~\eqref{im:4}, indicate that $\sS_d^{(\Sig)} (R)$  can be  interpreted as characterizing entanglement correlations {\it at scale $R$}. Thus in the continuum limit as we vary $R$ from zero to infinity, $\sSS$ can  be interpreted as describing the renormalization group (RG) flow of the ``renormalized entanglement entropy'' from short to large distances.  In contrast to the usual discussion of RG using some auxiliary mass or length scale, here we have the flow of a  physical observable with real physical distances.
Its derivative 
\be \label{imder}
 R {d \sSS \ov d R}
\ee
can then be interpreted as  the ``rate'' of the flow. With the usual intuition that RG flow leads to a loss of short-distance degrees of freedom, it is natural to wonder whether it also leads to a loss of entanglement. In other words, {\it could $\sSS$ also track the number of degrees of freedom of a system at scale $R$?} which would imply~\eqref{imder} should be negative, i.e. $\sSS$ should be monotonically decreasing.  

For $d=2$,\footnote{for which $\Sig$ is given by two points and there is no need to have a superscript in $\sS_2 (R)$.} a previous result of Casini and Huerta~\cite{Casini:2004bw} shows that 
 $\sS_2 (R)$ is indeed monotonically decreasing for all Lorentz-invariant, unitary QFTs, which  provides an alternative proof of Zamolodchikov's $c$-theorem~\cite{Zamolodchikov:1986gt}.
 
In higher dimensions, the shape of $\Sig$ also matters. We argue in Sec.~\ref{sec:dof}  that 
 $\sS_d^{(\rm sphere)} (R)$ has the best chance to be monotonic. At a fixed point, $\sS_d^{(\rm sphere)} (R)$ reduces to the previously proposed central charge in all dimensions.\footnote{That the entanglement entropy could provide a unified definition of central charge for all dimensions was recognized early on in~\cite{Ryu:2006ef} and was made more specific in~\cite{Myers:2010xs} including proof of a holographic $c$-theorem.}  Its monotonicity would then establish the conjectured $c$-theorems~\cite{Cardy:1988cwa,Myers:2010xs,Jafferis:2011zi} for each $d$.   
(For notational simplicity, from now on we will denote the corresponding quantities for a sphere simply as $S(R)$ and $\sS_d (R)$ without the superscript.)

 In Sec.~\ref{sec:free} we consider a free massive scalar and Dirac field in $d=3$, where available partial results again support that\footnote{A similar construction which involves the partition function (instead of entanglement entropy) has been used in~\cite{Friedan:2003yc} in connection with the g-theorem for $d=1$.}
\be
\label{c3}
  \sS_3 (R) =  R{\p S \ov \p R} - S 
  \ee 
is monotonic. 
 
In Sec.~\ref{sec:holo} and~\ref{sec:num}  we turn to holographic systems  whose gravity dual satisfies the {\it null-energy condition}. In Sec.~\ref{sec:holo}, among other things we 
show that when the central charges of the UV and IR fixed points are sufficiently close, $\sS_d (R)$ is always monotonic in all dimensions. Sec.~\ref{sec:num} is devoted to numerical studies of various holographic systems in $d=3$ and $d=4$. 
We find all the examples support the conjecture that: in $d=3$, {\it $\sS_3 (R)$ is always non-negative and monotonically decreasing with $R$ for  Lorentz-invariant, unitary QFTs.}  

In $d=4$, where 
\be
  \label{c4}
\sS_4 (R) =  {1 \ov 2} R\p_R (R \p_R S- 2 S) = {1 \ov 2} \le(
R^2 {\p^2 S \ov \p R^2} - R {\p S \ov \p R}\ri) \ 
\ee
we find that  while $\sS_4 (R)$ appears to have the tendency to be monotonically decreasing, there exist holographic systems where it, however, is not always monotonic and can become negative.

We conclude in Sec.~\ref{sec:conc} with a summary and a discussion of future directions. 

{\it Note Added:} When this paper is finalized, we became aware of~  \cite{Myers:2012ed} which has some overlap with our study. After the first version of this preprint appeared~\cite{Casini:2012ei} proved that $ \sS_3 (R)$ is indeed monotonic, thereby proving the $c$-theorem in three dimensions.

\section{A refinement of entanglement entropy} \label{app:UV}

In our discussion below we will assume that the system under consideration is equipped with a {\it bare} short-distance cutoff $\de_0$, which is much smaller than all other physical scales of the system. The continuum limit is obtained by taking $\de_0 \to 0$ while keeping other scales fixed.  The entanglement entropy for a spatial region is {\it not} a well-defined observable in the continuum limit as it diverges in the $\de_0 \to 0$ limit. The common practice is to subtract the UV divergent part by hand, a procedure which is often ambiguous. 
The goal of this section is to introduce a refinement of the entanglement entropy which is not only UV finite, but also is most sensitive to the entanglement correlations at the scale of the size of the entangled region.

\subsection{Structure of divergences in entanglement entropy} \label{sec:div}

In this subsection we consider the structure of divergent terms in the entanglement entropy. 
We assume that the theory lives in flat $\RR^{1,d-1}$ and is rotationally invariant.  
The discussion below is motivated from that in~\cite{tarun} which considers the general structure of local contributions to entanglement entropy in a gapped phase.\footnote{We thank Tarun Grover for  discussions.} We will mostly consider the vacuum state and will comment on the thermal (and finite chemical) state at the end.

Let us denote the divergent part of the entanglement entropy for a region enclosed by a surface $\Sig$
as $S_{\rm div}^{(\Sig)}$. Then $S_{\rm div}^{(\Sig)}$ should only depend on {\it local} physics at the cutoff scale near the entangling surface.  
For a smooth $\Sig$,  one then expects that $S_{\rm div}^{(\Sig)}$ should be expressible in terms of local geometric invariants of $\Sig$, i.e. 
\be \label{locE}
S_{\rm div}^{(\Sig)} = \int_\Sig {d^{d-2} \sig} \,  \sqrt{h} F (K_{ab}, h_{ab}) \
\ee
where $\sig$ denotes coordinates on $\Sig$, $F$ is a sum of all possible local geometric invariants formed from the induced metric $h_{ab}$ and extrinsic curvature $K_{ab}$ of $\Sig$. 
Note that here we are considering a surface embedded in flat space, all intrinsic curvatures and their derivatives can be expressed in terms of $K_{ab}$ and its tangential derivatives, thus all geometric invariants can be expressed in terms of  the extrinsic curvature and its tangential derivatives.
 The proposal~\eqref{locE} is natural
as $S_{\rm div}^{(\Sig)} $ should not depend on the spacetime geometry away from the surface nor how we parametrize the surface. Thus when the geometry is smooth, the right hand side is the only thing one could get after integrating out the short-distance degrees of freedom. In particular, the normal derivatives of $K_{ab}$ cannot appear as they depend on how we extend $\Sig$ into a family of surfaces, so is not intrinsically defined for the surface itself.

Here we are considering a pure spatial entangled region in a flat spacetime, for which the extrinsic curvature in the time direction is identically zero. Thus in~\eqref{locE}  we only have $K_{ab}$ for the spatial normal direction.  In more general situations, say if the region is not on a spatial hypersurface or in a more general spacetime,  then $\Sig$ should be considered as  a co-dimensional two surface in the full spacetime and in~\eqref{locE} we will have $K_{ab}^\al$ with $\al$ running over two normal directions.

Given~\eqref{locE}, now an important point is that in the vacuum (or any pure state), 
\be \label{pure}
S^{(\Sig)} = S^{(\bar \Sig)} 
\ee
where $S^{(\bar \Sig)} $ denotes the entanglement entropy for the region outside $\Sig$, and 
in particular 
\be \label{EqV}
S_{\rm div}^{(\Sig)} =  S_{\rm div}^{(\bar\Sig)}  \ . 
\ee
Recall that $K_{ab}$ is defined as the normal derivative of the induced metric and is odd under changing the orientation of  $\Sig$, i.e., in $S_{\rm div}^{(\bar\Sig)} $ it enters with an opposite sign. Thus~\eqref{locE} and~\eqref{EqV} imply that $F$ should be  an {\it even} function of $K_{ab}$. 
In a Lorentz invariant theory, there is also an alternative argument\footnote{We thank R. Myers for pointing this out to us.} which does not use~\eqref{pure} or~\eqref{EqV}. Consider a more general situation with both $K_{ab}^\al$ as mentioned above. The $\al$ index has to be contracted which implies that $F$ must be even in $K_{ab}^\al$. Then for a purely spatial surface we can just set  the time component of $K_{ab}^\al$ to zero, and $F$ is still even for the remaining $K_{ab}$.  
 
 As a result one can show that for a {\it smooth} and {\it scalable} surface $\Sig$ of size $R$, the divergent terms can only contain the following dependence on $R$ 
\be \label{divv}
S_{\rm div}^{(\Sig)} = a_1 R^{d-2} + a_2 R^{d-4} + \cdots \ .
\ee
See Appendix~\ref{app:ent} for a precise definition of scalable surfaces. Heuristically speaking, these are surfaces whose shape does not change with their size $R$, i.e. they are specified by a {\it single} dimensional parameter $R$ plus possible other dimensionless parameters describing the shape.  For such a surface, one can readily show that various quantities scale with $R$ as (see  Appendix~\ref{app:ent} for more details) 
\be \label{scalo}
h_{ab} \sim R^2, \qquad K_{ab} \sim R, \qquad D_a \sim R^0 
\ee
where $D_a$ denotes covariant derivative on the surface. As a result, any fully contracted quantity which is even in $K$, such as $F$ in~\eqref{locE}, can only give rise to terms proportional to $R^{-2n}$ with $n$ a non-negative integer, which then leads to~\eqref{divv}. Below we restrict our discussion to scalable  surfaces.

Now let us consider a scale invariant theory in the vacuum. On dimensional ground, the only other scale can appear in~\eqref{divv} is the short-distance cutoff $\de_0$. We should then have 
\be \label{dige}
a_1 \sim {1 \ov \de_0^{d-2}}, \qquad a_2 \sim {1 \ov \de_0^{d-4}}, \cdots 
\ee
and so on. For odd $d$, the $O(R^0)$ term is not among those in~\eqref{divv} and thus should be finite. 
For even $d$, there can be a $\log \de_0$ term at the order $O(R^0)$ and should come with $\log {R \ov \de_0}$ in order to have to the right dimension. We thus conclude that for a scale invariant theory, the entanglement entropy across a scalable surface $\Sig$ in the vacuum should have the form 
\be \label{enD}
S^{(\Sig)}= \bca {R^{d-2} \ov \de_0^{d-2}} + \cdots + {R \ov \de_0} +   (-1)^{d-1 \ov 2} s^{(\Sig)}_d
+ {\de_0 \ov R} + \cdots & {\rm odd \;\; d} \cr
{R^{d-2} \ov \de_0^{d-2}} + \cdots + {R^2 \ov \de_0^2} +  (-1)^{d -2 \ov 2}  s^{(\Sig)}_d  \log {R \ov \de_0} + {\rm const} + {\de_0^2 \ov R^2} + \cdots & {\rm even \;\; d} 
\eca
\ee
where for notational simplicity we have suppressed the coefficients 
of  non-universal terms. It is important to emphasize that $S$ does not contain any divergent terms with negative powers of $R$ in the limit $\de_0 \to 0$.
The form~\eqref{enD} was first predicted from holographic 
calculations in~\cite{Ryu:2006ef} for CFTs with a gravity dual.  $s^{(\Sig)}_d$ is an $R$-independent constant which gives the universal part of the entanglement entropy. The sign factors before $s^{(\Sig)}_d$ in~\eqref{enD} are chosen for later convenience. As indicated by the superscript, 
 $s^{(\Sig)}_d$ in general depends on the shape of the surface.

For a general QFT, there could be other mass scales, which we will denote collectively as $\mu$. Now the coefficients $a_i$ in~\eqref{divv} can also depend on $\mu$, e.g., we can write $a_1$ as 
\be \label{yuep}
a_1 = {1 \ov \de_0^{d-2}} h_1 (\mu \de_0) 
\ee
and similarly for other coefficients. Note that by definition of $\de_0$, we always have $\mu \de_0 \ll 1$ and $h_1$ can be expanded in a power series of $\mu \de_0$. Now for a 
renormalizable theory, the dependence on $\mu$ must come with a non-negative power, as when taking $\mu \de_0 \to 0$, $a_1$ should not be singular and should recover the behavior of the UV fixed point. 
In other words, for a renormalizable theory, the scale(s) $\mu$ arises from some relevant 
operator at the UV fixed point, which implies that $\mu \de_0$ should always come with a non-negative power in the limit $\mu \de_0 \to 0$. This implies that the UV divergences of $a_i$ should be no worse than those in~\eqref{dige}. 
In particular, there cannot be divergent terms 
with negative powers of $R$ for even $d$, and for odd $d$ the divergence should stop at order $O(R)$. 
These expectations will be confirmed by our study of holographic systems in Sec.~\ref{sec:asym} and~\ref{sec:pof1} (see e.g.~\eqref{bexp}), where we will find that $h_1 (\mu \de_0)$ has the expansion $h_1 (\mu \de_0)= c_0 + c_2  (\mu \de_0)^{2 \al} + c_3 (\mu \de_0)^{3 \al} + \cdots$, where $\al = d- \De$ with $\De$ the UV dimension of the leading relevant perturbation at the UV fixed point. 

So far  we have been considering  the vacuum. The discussion of the structure of divergences should work also for systems at finite temperature or finite chemical potential. In such a mixed state,  while~\eqref{pure} no longer holds, equation~\eqref{EqV} should still apply as the short-distance physics should be insensitive to the presence of temperature or chemical potential. Also recall that for a Lorentz invariant system, there is an alternative argument for~\eqref{divv} which does not use~\eqref{EqV}.

\subsection{Properties of $\sS^{(\Sig)} (R)$} \label{sec:prop}

Given the structure of divergent terms in $S^{(\Sig)} (R)$ discussed in the previous subsection, one 
can then readily check that when acting on $S^{(\Sig)} (R)$ with the differential operator in~\eqref{Scen}, all the UV divergent terms disappear and the resulting $\sS^{(\Sig)}_d (R)$ is finite in the continuum limit $\de_0 \to 0$. In fact, what the differential operator does is to eliminate any term (including finite ones) in $S^{(\Sig)} (R)$ which has the same $R$-dependence as the terms in~\eqref{divv}. We believe, for the purpose  of extracting long range correlations, it is sensible to also eliminate possible finite terms with the same $R$-dependence, as they are ``contaminated'' by short-distance correlations. 
In particular,  in the continuum limit this makes  $\sS^{(\Sig)}_d  (R)$ invariant under any redefinitions of the UV cutoff $\de_0$ which do not involve $R$.\footnote{Since $R$ is the scale at which we probe the system, reparameterizations of the short-distance cutoff should not involve $R$.} With a finite $\de_0$, $\sS^{(\Sig)}_d  (R)$ does depend on $\de_0$, but only very weakly, through inverses powers of ${\de_0 \ov R}$. This will be important in our discussion below. 
 
In the rest of this section we show that the resulting $\sS^{(\Sig)}_d (R)$ is not only UV finite, but also have various desirable features. In this subsection we discuss its behavior in the vacuum, while in Sec.~\ref{sec:ft} discuss its properties at a finite temperature and chemical potential.
 
 For a scale invariant theory, from~\eqref{enD} we find that for all $d$
 \be \label{ssce}
 \sS^{(\Sig)}_d (R) = s^{(\Sig)}_d
 \ee
is $R$-independent. The sign factors in~\eqref{enD} were chosen so that there is no sign factor
in~\eqref{ssce}.  Note that if we make a redefinition of the form $\de_0 \to \de_0  \le(1 + c_1 \mu \de_0 + c_2 (\mu \de_0)^2 + \cdots \ri)$ where $\mu$ is some mass scale, for odd $d$ the UV finite term in~\eqref{enD} is modified. But $\sS^{(\Sig)}_d (R)$, and  $s^{(\Sig)}_d$ as defined from~\eqref{ssce}, is independent of this redefinition.

Let us now look at properties of $\sS^{(\Sig)}_d (R)$ for a general renormalizable QFT (i.e. with a well-defined UV fixed point). 
Below we will find it convenient to introduce a floating cutoff $\de$, which we can adjust depending on scales of interests. 
At the new cutoff $\de$, the system is described by the Wilsonian effective action $I_{eff} (\de;\de_0)$, which is obtained by integrating out degrees of freedom from the bare cutoff $\de_0$ to $\de$. 
The entanglement entropy $S^{(\Sig)} (R; \de_0, \de)$ calculated from $I_{eff} (\de;\de_0)$ with cutoff $\de$ should be 
independent of choice of $\de$. So should the resulting $\sS^{(\Sig)}_d (R)$. Below we will consider the continuum limit, i.e. with bare cutoff  $\de_0 \to 0$. 

First consider the small $R$ limit, i.e. $R$ is much smaller than any other length scale of the system. 
Clearly as $R \to 0$, these other scales should not affect $S^{(\Sig)} (R)$, which should be given by its expression at the UV fixed point. Accordingly, $\sS^{(\Sig)} _d (R)$ also reduces to that of the UV fixed point, i.e. 
\be \label{irim}
\sS_d^{(\Sig)} (R)  \to s^{(\Sig, \rm UV)}_d, \quad R \to 0 \ .
\ee
As we will see in Sec.~\ref{sec:Rasy}, 
studies of holographic systems (with $\Sig$ given by a sphere) predict that the leading small $R$ correction to~\eqref{irim} is given by
\be \label{smmr}
\sS_d^{(\Sig)} (R)  =  s^{(\Sig, \rm UV)}_d  + O((\mu R)^{2 \al})  \quad R \to 0 \ 
\ee
where $\al = d - \De$ with $\De < d$ the UV dimension of the leading relevant scalar perturbation. 
Equation~\eqref{smmr} has a simple interpretation that the leading contribution from a relevant operator comes at two-point level. We believe it can be derived in general, but will not pursue it here.

The story is more tricky in the large $R$ limit, as all degrees of freedom at scales between the bare UV cutoff $\de_0$ and $R$ could contribute to the entanglement entropy $S^{(\Sig)} (R)$ 
in this regime. Nevertheless, one can argue that 
\be \label{lrli}
\sS_d^{(\Sig)} (R)  \to s_d^{\rm(\Sig, IR)}, \quad R \to \infty 
\ee
as follows. When $R$ becomes much larger than all other length scales of the system, we can choose a floating cutoff $\de$ 
to be also much larger than all length scales of the system while still much smaller than $R$, i.e. 
\be 
{1 \ov \mu_1}, {1 \ov \mu_2}, \cdots \ll \de \ll R \
\ee
where $\mu_i, i=1,2,\cdots$ denote possible mass parameters of the system.  Now the physics between $\de$ and $R$ is controlled by the IR 
fixed point, i.e. we should be able to write $S^{(\Sig)} (R)$ again as~\eqref{enD}, but with $\de_0$ replaced by $\de$, and $s_d^{(\Sig)} $ by  $s_d^{\rm(\Sig, IR)}$. Then equation~\eqref{lrli} immediately follows. 
In other words, while in terms of the bare cutoff $\de_0$, the entanglement entropy $S^{(\Sig)}  (R) = S^{(\Sig)}  (\de_0, R, \mu_1, \mu_2, \cdots)$
could be very complicated in the large $R$ regime, involving many different scales, there must exist a redefinition of short-distance cutoff $\de_{\rm IR} = \de_{\rm IR} (\de, \mu_1, \mu_2, \cdots)$, in terms of which $S^{(\Sig)} (R)$ reduces to
the standard form~\eqref{enD} with $\de$ replaced by $\de_{\rm IR}$, and $s_d^{(\Sig)} $ by  $s_d^{\rm(\Sig, IR)}$. In fact, higher order terms in~\eqref{enD} with negative powers of $R$ also imply that generically we should expect the leading large $R$ corrections to~\eqref{lrli} to have the form
\be \label{larr}
\sS_d^{(\Sig)} (R)  = s_d^{\rm(\Sig, IR)} + \bca O({1 \ov R}) & {\rm odd \; d} \cr
                                                              O({1 \ov R^2}) &  {\rm even \; d}
                                                              \eca
    , \quad R \to \infty \ .
\ee
This expectation is supported by theories of  free massive scalar and Dirac fields as we will see in Sec.~\ref{sec:free}, and by holographic systems as we will see in Sec.~\ref{sec:Rasy}.
Holographic systems also predict an exception to~\eqref{larr} which happens when the flow away from the IR fixed point toward UV is generated by an irrelevant operator with IR dimension $ \De_{\rm IR}$ sufficiently close to $d$, for which we have instead  (see Sec.~\ref{sec:Rasy})\footnote{The expression below is derived in Sec.~\ref{sec:Rasy} for $\Sig$ given by a sphere and closely separated UV/IR fixed points. We believe the result should be more general, applicable to generic systems and smooth $\Sig$, but will not pursue a general proof here.} 
\be \label{larr1}
\sS_d^{(\Sig)} (R)  = s_d^{\rm(\Sig, IR)} + O \le({1 \ov (\tilde \mu R)^{2 \tilde \al}} \ri), \quad 
{\rm for} \;  \bca \tilde \al < \ha & {\rm odd \; d} \cr
                                    \tilde \al     <  1 &  {\rm even \; d}
                                                              \eca
    , \quad R \to \infty \ ,
\ee
where $ \tilde \al = \De_{\rm IR} - d $. 

By adjusting the floating cutoff $\de$, one can also argue that $\sS_d^{(\Sig)}  (R)$ should be most sensitive to contributions from degrees of freedom around $R$. 
Consider e.g. a length scale $L_1$ which is much smaller than $R$. 
In computing $S^{(\Sig)}  (R)$, we can choose  a floating short-distance cutoff $ \de$ which satisfies 
\be 
L_1 \ll  \de \ll R \ .
\ee
As discussed at the beginning of this subsection, by design $\sS_d^{(\Sig)}  (R)$ is insensitive to short-distance cutoff $\de$  when $\de \ll R$.\footnote{Of course ultimately as mentioned earlier $\sS_d^{(\Sig)}  (R)$ should be independent of choice $\de$, when one includes all possible dependence on $\de$ including those in coupling constants. Here we are emphasizing that even explicit dependence on $\de$ should be suppressed by negative powers of ${\de \ov R}$.} 
We thus conclude that $\sS_d^{(\Sig)}  (R)$  should be insensitive to contributions of from d.o.f around $L_1$. 

While our above discussion around and after~\eqref{lrli}  assumes a conformal IR fixed point, the discussion also applies to when the IR fixed point is a gapped phase, where there are some  differences depending on the spacetime dimension. For  odd $d$, using $d=3$ as an illustration, the entanglement entropy for a smooth surface $\Sig$ in a gapped phase  has the form (see e.g. also~\cite{tarun})
\be 
S^{(\Sig)} (R) = \al R -  \ga + O(R^{-1})
\ee
where $\ga$ is the topological entanglement entropy~\cite{wenx,Kita}.   We then have
\be 
\sS_3^{(\Sig)}  (R) \to \ga, \qquad R \to \infty \ .
\ee
In gapped phases without topological order, $\ga =0$.
Thus a nonzero $\sS_3^{(\Sig)} (R \to \infty)$ signals the system has long range entanglement, i.e. the system is either gapless or topological-ordered in the IR. The two cases can be distinguished in that for a topological ordered phase $\ga$ should be shape-independent, but in a gapless case, $s_3^{(\rm \Sig,IR)}$ in~\eqref{irim} is shape-dependent.

For even $d$, in a gapped phase we expect that $S^{(\Sig)} (R)$ does not have a term proportional to $\log R$ for large $R$, and thus we should have 
\be 
\sS_{2n}^{(\Sig)} (R) \to 0, \quad R \to \infty , \qquad n=1,2,\cdots \ .
\ee
Nevertheless, it has been argued in~\cite{tarun} that the size-independent part of the entanglement entropy contains 
topological entanglement entropy. 
Such a topological term could not be captured by $\sS_d^{(\Sig)}  (R)$, as all terms in~\eqref{Scen} contain derivatives with respect to $R$ for even $d$.  This is not surprising, as in even $d$, the $R$-independent part of the entanglement entropy also contains a finite non-universal local part, as is clear from the discussion around~\eqref{divv}. Thus it is not possible to separate the topological from the non-universal contribution using a single connected entangling surface, and one has to resort to constructions like those in~\cite{wenx,Kita} to consider combination of certain regions in such a way that the local part cancels while the topological part remains~\cite{tarun}.

\subsection{Finite temperature and chemical potential} \label{sec:ft}

As discussed at the end of Sec.~\ref{sec:div}, we expect $\sS_d^{(\Sig)} (R)$ should also be UV finite in the continuum limit at a finite temperature or chemical potential. Here we briefly discuss its properties, and for simplicity will restrict to a scale invariant theory.\footnote{See also~\cite{sw2} regarding scaling behavior of the entanglement entropy at finite $T$. In ref.~\cite{sw2} considered the entanglement entropy itself with UV part subtracted manually. 
}

For a scale invariant system at a finite temperature $T$, since there is no other scale in the system, $\sS_d^{(\Sig)} (R, T)$ must have a scaling form, i.e. 
\be \label{tscal}
 \sS_d^{(\Sig)} (R, T) = \sS_d^{(\Sig)} (RT) \ .
 \ee
 In particular, in the high temperature limit, i.e. $RT \gg 1$ it must be dominated by thermal entropy at leading order, while in the low temperature limit $RT \to 0$, it should reduce to $s_d^{(\Sig)}$. 
  For a scale invariant theory, the thermal entropy has the form $S^{\rm Th}_d = \eta_d T^{d-1} V_{\Sig} \sim (T R)^{d-1}$, where $V_\Sig$ is volume of the spatial region enclosed by $\Sig$ and $\eta_d$ is some constant. 
Thus we should have  
  \be
  \sS_d^{(\Sig)} (RT) \to \bca  S^{\rm Th}_d  & {\rm odd \; d} \cr
                             {(d-1)!! \ov (d-2)!!} S^{\rm Th}_d  & {\rm even \; d}
  \eca  , \quad R T \to \infty \ .
  \ee
More explicitly, for $d=3$ we expect when $RT \gg 1$,   
  \be 
  \sS_d^{(\Sig)} (RT)  = \eta_d T^2 V_{\Sig} + c_d^{(\Sig)} + \cdots 
  \ee 
where the first term is simply the thermal entropy, $\cdots$  denotes terms with negative powers of $RT$. The second term $c_d^{(\Sig)}$ is a constant.   
It would be interesting to compute this constant for some explicitly examples to see whether some physical interpretation (or significance) can be attached  to it.   

Similarly, with a nonzero chemical potential $\mu$, as a generalization of~\eqref{tscal} we expect that
\be \label{muscal}
\sS_d^{(\Sig)} (R, \mu, T) = \sS_d^{(\Sig)} \le({T \ov \mu}, RT \ri)   \ .
 \ee

\section{Entanglement entropy of a (non)-Fermi liquid}  \label{sec:fermi}

In this section we show that the entanglement entropy of a (non)-Fermi liquids can be obtained by simple dimensional analysis.\footnote{See also a recent discussion in~\cite{sw2} based on finite temperature scaling and crossover. See also~\cite{Ogawa:2011bz,lizaetal,Shaghoulian:2011aa,Iizuka:2012iv} for recent discussion of logarithmic enhancement of holographic ``non-Fermi liquids.''}  Consider a $d$-dimensional system of a finite fermions density whose ground state is described by a Fermi surface of radius $k_F$. We have in mind a Fermi liquid, or a non-Fermi liquid described by the Fermi surface coupled to some gapless bosons (as e.g. in~\cite{sungsik}).  In either case, the low energy dynamics of the system involves fermionic excitations locally in momentum space near the Fermi surface, and different patches of the Fermi surface whose velocities are not parallel or anti-parallel to each other essentially decouple. In particular, $k_F$ drops out of the low energy effective action. We thus expect in the large $R$ limit, 
the ``renormalized entanglement entropy'' $\sSS$ should be proportional to the area of the Fermi surface $A_{FS} \propto k_F^{d-2}$, which can be considered as the ``volume'' of the available phase space.  Since there is no other scale in the system than $R$, $\sSS$ should then have the form 
\be \label{area}
\sSS \propto k_F^{d-2} R^{d-2}  \propto A_{FS} A_{\Sig} , \qquad R \to \infty \ .
\ee
where $A_\Sig$ denotes the area of the entangling surface $\Sig$. 
In other words, our ``renormalized entanglement entropy'' should satisfy a ``area law.''  Using~\eqref{Scen} one can readily see that the area law~\eqref{area} translates into the well-known behavior in the original entanglement entropy~\cite{FS1,FS2} (see also~\cite{sw1})
\be 
 S^{\Sig} (R) \propto  
 k_F^{d-2} R^{d-2} \log (k_F R) + \cdots \propto A_{FS} A_{\Sig} \log (A_{FS} A_{\Sig} ) + \cdots, 
  \ee 
where $k_F$ in the logarithm is added on dimensional ground and $\cdots$ denotes other non-universal parts. We note that this result does not depend on whether the Fermi surface has quasi-particles or not, i.e. whether it is a Fermi or non-Fermi liquid, only depends on the expectation that $\sSS$ is proportional to the area of the Fermi surface.  

This analysis can also be immediately generalized to predict the qualitative behavior of the entanglement entropy of higher co-dimensional Fermi surfaces. For a co-dimensional $n$ Fermi surface we should have\footnote{We define the co-dimension with respect to the full spacetime dimension $d$.} 
\be 
\sSS  \propto (k_F R)^{d-n} 
\ee
which implies that in the entanglement entropy itself\footnote{These results were also obtained by B.~Swingle (unpublished).}
\be \label{ferb}
S^{(\Sig)} (R) \propto \bca (k_F R)^{d-n} \log (k_F R)  & {\rm n \; even} \cr
                                   (k_F R)^{d-n}  & {\rm n \; odd} 
\eca \ .
\ee
Thus we find that there is a $\log R$ factor only for even co-dimensional Fermi surfaces. These results are again independent of whether there are quasi-particles.  Note that for a Fermi point where $n=d$, equation~\eqref{ferb} in consistent with one's expectation that for massless fermions there is a universal $\log R$ term only for even $d$. 
For general $n$, at least for free fermions, the alternating behavior of logarithmic enhancement in~\eqref{ferb} may also be understood (by generalizing an argument of~\cite{sw1}) as follows: 
at each point of a co-dimensional $n$ Fermi surface, there is an $n$-dimensional free fermion CFT. The $\log R$ appearance in~\eqref{ferb} is then consistent with  the fact that for an $n$-dimensional CFT, there is a universal $\log R$ piece only for $n$ even. 

It would be interesting to see whether our discussion may also be used to understand the logarithmic enhancement in the entanglement entropy of the critical spin liquids in~\cite{zhang} which are described by a projected Fermi sea state. 

\section{Renormalized  R\'enyi entropies}

In addition to the entanglement entropy, other important measures of entanglement properties of quantum states include R\'enyi entropies, which are defined as
\be
 S^{(\Sig)}_n \equiv -{1\ov n-1}\, \log \Tr \rho^n \ .
\ee
The entanglement entropy can be obtained from them by analytic continuation in $n$:
\be
\lim_{n\to 1} S^{(\Sig)}_n=S^{(\Sig)} \ .
\ee

The discussions of Sec.~\ref{app:UV}--Sec.~\ref{sec:fermi} for the entanglement entropy can be applied 
almost without any change to R\'enyi entropies. The main results include: 

\ben 

\item The divergent pieces of $S^{(\Sig)}_n$ should be expressible in terms of the local geometric invariants 
as in~\eqref{locE}. 
 
\item For a pure state,  equations \eqref{pure}--\eqref{EqV} apply to R\'enyi entropies. As a result the renormalized  R\'enyi entropies $\sS^{(\Sig)}_{n,d}$, obtained by acting the differential operators in~\eqref{Scen}
 to $S^{(\Sig)}_n$, are UV finite.  

\item For a CFT, the R\'enyi entropies $S^{(\Sig)}_n$ have the same structure as~\eqref{enD}, i.e.  
\be \label{renD}
S^{(\Sig)}_n (R) = \bca {R^{d-2} \ov \de_0^{d-2}} + \cdots + {R \ov \de_0} +   (-1)^{d-1 \ov 2} s^{(\Sig)}_{n,d}
+ {\de_0 \ov R} + \cdots & {\rm odd \;\; d} \cr
{R^{d-2} \ov \de_0^{d-2}} + \cdots + {R^2 \ov \de_0^2} +  (-1)^{d -2 \ov 2}  s^{(\Sig)}_{n,d}  \log {R \ov \de_0} + {\rm const} + {\de_0^2 \ov R^2} + \cdots & {\rm even \;\; d} 
\eca
\ee
with 
\be 
\sS^{(\Sig)}_{n,d} (R) = s^{(\Sig)}_{n,d} = {\rm const} \ .
\ee

\item For a general (renormalizable) QFT $\sS^{(\Sig)}_{n,d} (R)$ interpolate between the values of the UV and IR fixed point 
\be \label{asumRe}
\sS^{(\Sig)}_{n,d} (R) \to \bca  s_{n,d}^{(\rm \Sig, UV)} & R \to 0 \cr
                  s_{n,d}^{({\rm \Sig, IR})} & R \to \infty 
                  \eca \ ,
\ee
and are most sensitive to the degrees of freedom at the scale $R$.

\item For a scale invariant theory at finite temperature and chemical potential,  $\sS^{(\Sig)}_{n,d}$ should take the scaling form 
\be \label{muscalR}
\sS_{n,d}^{(\Sig)} (R, \mu, T) = \sS_{n,d}^{(\Sig)} \le({T \ov \mu}, RT \ri)   \ .
 \ee
Unlike for entanglement entropy we do not expect a simple relation with the thermal entropy in the high temperature limit.  

\item All R\'enyi entropies contain logarithmic violations of the area law for a (non)-Fermi liquid
\be 
 S^{(\Sig)}_n (R) \propto A_{FS} A_{\Sig} \log (A_{FS} A_{\Sig} ) + \cdots .
  \ee 
This generalizes a previous result for the free Fermi gas~\cite{calaetal}.

\een

The key difference between entanglement entropy and the R\'enyi entropies is that strong subadditivity does not hold for the latter. In the following sections we discuss how entanglement entropy is related to the number of degrees of freedom. These relations do not appear to have obvious generalization to R\'enyi entropies. 


\section{Entanglement entropy as measure of number of degrees of freedom} \label{sec:dof}

For the rest of this paper we will restrict our discussion to the renormalized entanglement entropy in the vacuum. In Sec.~\ref{app:UV} we showed that in the vacuum $\sSS$ introduced in~\eqref{Scen} has various desirable features:
 
\ben 

\item  It  has a well-defined continuum limit. 

\item For a CFT, it is independent of $R$ and given by the universal part of the entanglement entropy~\eqref{enD}
\be \label{sjep}
\sSS = \sds \ .
\ee 

\item For a renormalizable quantum field theory, it 
interpolates between the values of UV and IR fixed points as $R$ is increased from zero to infinity.  
\be \label{asum}
\sSS \to \bca  s_d^{(\rm \Sig, UV)} & R \to 0 \cr
                  s_d^{({\rm \Sig, IR})} & R \to \infty 
                  \eca \ .
\ee
It should be understood that in~\eqref{asum} if the IR fixed point is described by a gapped phase, then 
$s_d^{({\rm IR})}$ is either given by the topological entanglement entropy (for odd $d$) or zero (for even $d$).

\item It is most sensitive to degrees of freedom at  scale $R$.

\een 
Thus $\sSS$  provides a set of observables which can be used to directly probe and characterize quantum entanglement at a given scale $R$. As discussed in the Introduction, these properties also imply that we may interpret the dependence on $R$ as a RG flow. 
A natural question which then arises is whether $\sSS$ could also provide a scale-dependent measure of the number of degrees of freedom of a system. Given the physical intuition that RG leads to a loss of degrees of freedom, a necessary condition for this interpretation is then
\be \label{monb}
R {d \sSS \ov d R} < 0 \ 
\ee
which in turn requires (given~\eqref{asum}) 
\be \label{ccond}
s_d^{(\rm \Sig, UV)} > s_d^{(\rm \Sig, IR)} \ .
\ee   
Note that~\eqref{ccond} alone is enough to establish $s_d^{(\Sig)}$ as a measure of the number of degrees of freedom for CFTs, while establishing $\sSS$ as a measure of degrees of freedom for general QFTs requires a much stronger condition~\eqref{monb}.

 For $d=2$, the entangled region becomes an interval (there is no shape difference) and equation~\eqref{Scen} reduces to\footnote{The function~\eqref{2d} has also been discussed e.g. in~\cite{Casini:2009sr,Hertzberg:2010uv} as the universal part of the entanglement entropy in $d=2$.}  
\be \label{2d}
 \sS_2 (R) = {R} {d S \ov dR} \ ,
 \ee
which for a CFT then gives~\cite{Callan:1994py,Holzhey:1994we,Calabrese:2004eu}
\be
\sS_2 = {c \ov 3}
\ee
where $c$ is the central charge. In this case, Zamolodchikov's $c$-theorem~\cite{Zamolodchikov:1986gt} ensures~\eqref{ccond} and there 
 exists a  beautiful proof by Casini and Huerta~\cite{Casini:2004bw}
showing that $\sS_2 (R)$ is indeed monotonically decreasing for all Lorentz-invariant, unitary QFTs.  
 Note that while there already exist an infinite number of 
 $c$-functions~\cite{Cappelli:1990yc} including Zamolodchikov's original one, $\sS_2 (R)$ has some special appeal, given that it also characterizes the entanglement of a system. We would like to propose that it gives a ``preferred'' $c$-function which best characterizes the number of d.o.f. of a system at scale $R$.

In higher dimensions, the shape of $\Sig$ also matters. Could~\eqref{monb} and~\eqref{ccond}  apply to generic or only certain shapes? For this purpose, consider first the weaker condition~\eqref{ccond}.

 For even $d \geq 4$, since $\sds$ appears as the coefficient of the divergent term $\log \de_0$ in~\eqref{enD}, it can be expressed in terms of integrals of the geometric invariants associated with $\Sig$~(recall~\eqref{locE}), and in particular related to trace anomaly~\cite{Ryu:2006ef,solodu}. For example, for $d=4$~\cite{Ryu:2006ef,solodu}
\be \label{genes}
s_4^{(\Sig)} = {2 a_4} \int_\Sig d^2 \sig \sqrt{h} \, E_2 + {c_4} \int_\Sig d^2 \sig \sqrt{h} \, I_2
\ee
where $a_4$ and $c_4$ are coefficients of the trace anomaly\footnote{We use the convention
 \be
\vev{T^a{}_a} =-\frac{c_4}{16\pi^2}
W_{abcd}W^{abcd}+  2 a_4 E_4
 \ee
 with $W_{abcd}$ the Weyl tensor and $E_4 = \frac{1}{32\pi^2}\left( R_{abcd}R^{abcd}-4R_{ab}R^{ab}
+ R^2\right)$ the Euler density. The relation to $A$ and $B$ of~\cite{solodu} is $A = {4 a_4 \ov \pi^2}$, $B = {4 c_4 \ov \pi^2}$. Also note there is a minus sign in the definition of $s_4^{(\Sig)}$ in~\eqref{enD}. In comparing with~\cite{solodu} we have also set the extrinsic curvature in the time direction to zero.} and (below $K= K_{ab} h^{ab}$)
 \bea \label{euler}
 E_2 &=& {1 \ov 4 \pi} 
  \sR = {1 \ov 4 \pi}   \le(K^2 -  K_{ab} K^{ab} \ri) \\
  I_2 &= & {1 \ov 2 \pi}   \le( \ha K^2  -  K_{ab} K^{ab} \ri)  \ .
 \label{weyl}
 \eea 
In~\eqref{euler} $\sR$ is the intrinsic curvature on $\Sig$ and in the second equality of~\eqref{euler} we have used the Gauss-Codacci relation in flat space. $E_2 $ is the Euler density for $d=2$ and $\int_\Sig d^2 \sig \sqrt{h} E_2$ is a topological invariant with value $2$ for a surface with spherical topology.  
 $I_2$ is a Weyl invariant and is zero for a sphere.  For a sphere we then have 
\be \label{anoi}
s_4^{(\rm sphere)} =  4  a_4 \ 
\ee 
while for other shapes, $s_4^{(\Sig)}$ will be a linear combination of $a_4$ and $c_4$. More than twenty years ago,  Cardy conjectured that~\cite{Cardy:1988cwa} $a_4$ and its higher dimensional generalizations obey the analogue of $c$-theorem. Only very recently was it proven for $d=4$~\cite{Komargodski:2011vj}. 
In addition, there are strong indications any combination of $a_4$ and $c_4$ (including $c_4$) will not  
satisfy such a condition~\cite{Komargodski:2011vj}.
Thus for $d=4$ {\it only} for $\Sig = {\rm sphere}$ could the condition~\eqref{ccond} be satisfied. 
For higher even dimensions the situation is less clear, but one again has~\cite{Ryu:2006ef,Myers:2010xs,Casini:2011kv}\footnote{In addition,  the structure of~\eqref{genes} persists in higher even dimensions except that there are more Weyl invariants~\cite{Myers:2010xs,hung2}.}
\be 
s_d^{(\rm sphere)} =  4  a_d \ .
\ee 
 

For odd $d$, $\sds$ does not arise from local terms in~\eqref{enD}, thus we do not expect that it can be expressed in terms of local geometric invariants on $\Sig$. This is in contrast to the local shape dependence in the even dimensional case. It would be interesting to understand how $\sds$ depends on the shape of the entangling surface. It was found in~\cite{Casini:2011kv} that for a sphere  
 \be \label{rel2}
s_d^{(\rm sphere)} = (\log Z)_{\rm finite}
\ee
where $ (\log Z)_{\rm finite}$ is the finite part of the Euclidean partition for the CFT on $S^d$.  Some support has been found that $s_d^{(\rm sphere)}$ (equivalently $(\log Z)_{\rm finite}$) satisfies the condition~\eqref{ccond}~\cite{Myers:2010xs,Jafferis:2011zi}.


To summarize, for $\Sig$ given by a sphere, there are (strong) indications that $s_d^{(\rm sphere)}$ could satisfy~\eqref{ccond} and thus provide a measure of the number of degrees of freedom for CFTs in all dimension~(including both odd and even)~\cite{Myers:2010xs,Casini:2011kv}. 
Below we will simply call $s_d^{(\rm sphere)}$ the central charge of a CFT. For $d \neq 4$, other shapes could still provide a similar measure, which will be left for future investigation. 
For the rest of the paper we study the stronger condition~\eqref{monb} for  $\Sig = {\rm sphere}$ in $d=3$ and $d=4$.
For notational simplicity,  we will drop the superscript ``sphere'' in various quantities and denote them simply
as $S(R), \sS_d (R)$ and $s_d$. 





\section{Free massive scalar and Dirac fermions in $d=3$} \label{sec:free}


In this section we consider $\sS_3 (R)$ for a free massive scalar and Dirac field. 
 For a free massive field, we expect that $\sS_3 (R)$  should approach that for a massless field as $R \to 0$, and $0$ as $R \to \infty$. We would like to see whether it is monotonic and positive in between. 
Recently, in the limit of $m R \gg 1$, it was found in~\cite{Hertzberg:2010uv,Huerta:2011qi} that ($\de$ is a short-distance cutoff)
\bea \label{sc1}
S_{\rm scalar}(mR)&=&\#{R \ov \de} -  {\pi \ov 6} \, mR  - {\pi \ov 240}\  {1 \ov mR} + \cdots 
\\
\label{fer1}
S_{\rm Dirac}(mR)&=&\#{R \ov \de} -  {\pi \ov 6} \, mR  - {\pi \ov 40}\  {1 \ov mR} + \cdots \ .  
\eea
From~\eqref{c3} we thus find that  
\bea \label{scle}
\sS_{\rm scalar}(mR)&=& + {\pi \ov 120}\  {1 \ov mR} + \cdots 
\\
\sS_{\rm Dirac}(mR)&=& + {\pi \ov 20}\  {1 \ov mR} + \cdots  
\eea
which are indeed monotonically decreasing with $R$. Note that the ${1 \ov mR}$ fall-off in the above expressions is also consistent with our earlier expectation~\eqref{larr}. 

We emphasize that if one simply subtracts the divergent part in~\eqref{sc1}, then 
the resulting 
\be
S^{\rm (finite)}_{\rm scalar} (mR) =-  {\pi \ov 6} \, mR  - {\pi \ov 240}\  {1 \ov mR} + \cdots 
\ee
 approaches minus infinity linearly as $R \to \infty$ and thus does {\it not} have a good asymptotic limit.
The presence of such a linear term can be understood as a finite renormalization between the short distance cutoffs of the UV and IR fixed points, as discussed in Sec.~\ref{app:UV}. 
  In contrast, $\sS_3 (R)$ approaches zero as $R \to \infty$ 
as one would expect of a system with a mass gap.

We have also calculated $\sS_3 (R)$ numerically for a massive scalar field\footnote{Compared to that of a scalar, the computation for a Dirac fermion requires significantly more computer time to achieve the same accuracy. We will leave it for future investigation.}  for a range of $mR$ 
as shown in Fig.~\ref{fig:frm} (see Appendix~\ref{app:num} for details for the numerical calculation).
The numerical result is consistent with our expectation of the limiting values of $\sS_3 (R)$ in the small and large $R$ limits, and also suggests that it is monotonic in between.

\begin{figure}[!h]
\begin{center}
\includegraphics[scale=0.5]{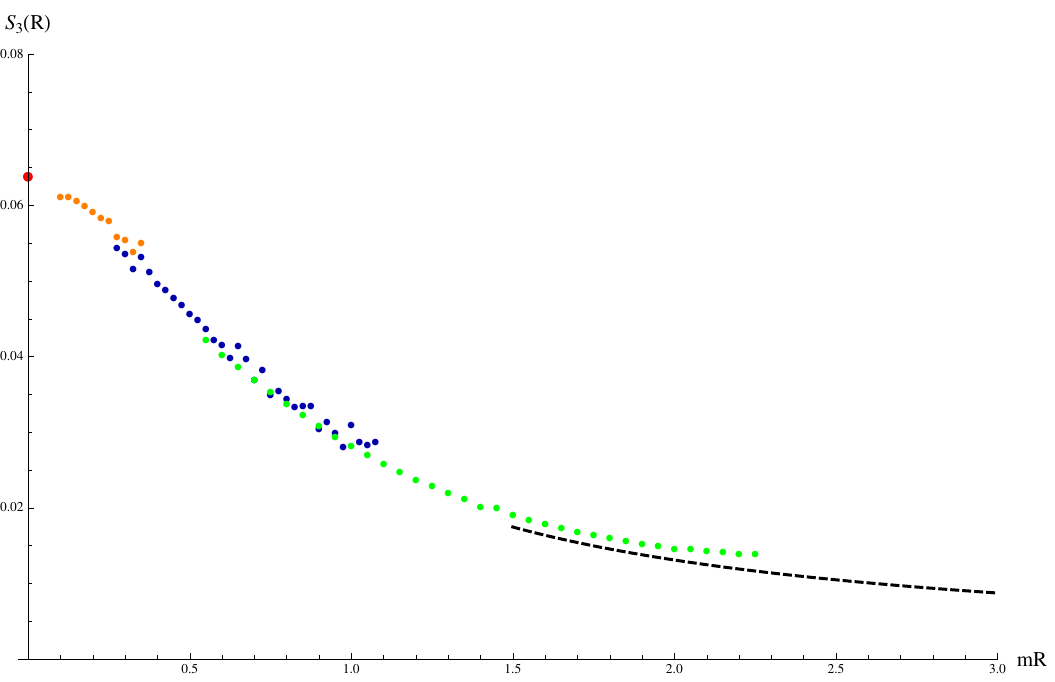}
\end{center}
\caption{$\sS_3(R)$ for a free massive scalar field: The red point is the value for $m=0$. The black dashed line is the result of the asymptotic expansion~\eqref{scle}.
The numerical results are computed following the method of~\cite{Srednicki:1993im} with a radial lattice discretization. We choose the system size to be $L_{IR}=200 a$, where $a$ is the lattice spacing. To avoid boundary effects the restriction to $10a \leq R \leq 45 a$ was made. To extend the range of $mR$ we obtained the results for $1/m=20 a, \ 40a, \ 120a$. In the plots, the orange dots are data points  for  $1/m= 120a$, the blue ones are for $1/m=40a$, and the green ones are for  $1/m=20 a$.
As expected all our data points collapse into one curve as $\sS_3(R)$  can only depend on $mR$ in the continuum limit.  For more details see Appendix~\ref{app:num}.
}
\label{fig:frm}
\end{figure}

\section{$\sS_d (R)$ for Holographic flow systems} \label{sec:holo}

In this section we discuss properties of $\sS_d (R)$ (defined for a sphere) for systems with a gravity dual using the proposal of~\cite{Ryu:2006bv,Ryu:2006ef} (see~\cite{Nishioka:2009un} for a review), which relates the entanglement entropy to the area of a minimal surface. Other recent discussion of entanglement entropy in holographic RG flow systems include~\cite{Myers:2010xs,deboer,jimliu,sinha,Paulos,Albash:2011nq}

 We will restrict our discussion to $d \geq 3$. After a brief discussion of the general set-up, we derive a relation between $\sS_d$ and the undetermined constant in the asymptotic expansion of the minimal surface. We then show that when the central charges of the IR and UV fixed points are close, for all dimensions, $\sS_d$ is always  monotonically deceasing with $R$ at leading order in the expansion ot difference of central charges of the UV and IR fixed points. Thus for flows between two sufficiently closely  separated fixed points, $\sS (R)$ appears to provide a good central function.

\subsection{Gravity set-up}

We consider a bulk spacetime which describes a renormalization group flow in the boundary theory. 
We assume that the system is Lorentz invariant. The flow can be induced either by turning on the source for some relevant scalar operators or by certain scalar operators developing vacuum expectation values (without a source). Below we will refer to them as source and vev deformation respectively. We denote the corresponding bulk fields by $\phi_a$. 

The bulk action can be written as 
\be 
I = {1 \ov 2 \ka^2} \int d^{d+1} x \, \sqrt{-g} \le[R - \ha G_{ab} \p \phi_a \p \phi_b - V (\phi_a) \ri]
\ee
where $G_{ab}$ is some positive-definite metric on the space of scalar fields. The spacetime metric
can be written in a form 
\be \label{1}
ds^2  =  {L^2 \ov z^2} \le(- dt^2 + d \rho^2 + \rho^2 d \Om_{d-2}^2 + {dz^2 \ov f(z)} \ri) \ .
\ee
We assume that $V(\phi_a)$ has a critical point at $\phi_a =0$ with $V(0) = -{d (d-1) \ov L^2}$, which corresponds to the UV fixed point. Near the boundary $z =0$, 
\be 
\phi_a (z) \to 0, \quad f (z) \to  1, \qquad z \to 0
\ee 
and the spacetime metric is that of AdS$_{d+1}$ with curvature radius $L$. 
Einstein equations and positive-definiteness of the kinetic term coefficients $G_{ab}$ require that the evolution of $f$ with $z$ should satisfy~\cite{Freedman:1999gp} 
\be \label{pep2}
\p_z f (z) > 0 \ 
\ee
i.e. $f(z)$ is a monotonically increasing function. More generally, equation~\eqref{pep2} follows from the null energy condition and Einstein equations, regardless of specific  form of the scalar action. 

At small $z$, $f(z)$ can be expanded as 
\be  \label{find}
f (z) = 1 + \mu^{2 \al} z^{2 \al}  + \cdots , \qquad z \to 0
\ee
where $\mu$ is some mass scale and $\al$ some positive constant.  For a source deformation,  $\al = d - \De$  with $\De$ the UV dimension of the {\it leading} relevant perturbating operator (i.e. the one with the smallest $\al$).\footnote{Note that when $\De = {d \ov 2}$ for which $\al = {d \ov 2}$, we should replace the second term in~\eqref{find} by $(\mu z)^d (\log \mu z)^2$.}  For a vev deformation we have $\al = \De$.\footnote{The above description is for the standard quantization. In the alternative quantization (which applies to ${d \ov 2} - 1 < \De < {d \ov 2}$),  we have instead $\al = \De$ for a source deformation and $\al = d - \De$ for  a vev deformation.}

As $z \to \infty$, we can have the following two possibilities: 

\ben 

\item  Flow to an IR conformal fixed point. In this case $\phi_a$ approaches a neighboring critical point 
$\phi_a = \phi^{(0)}_a$ with $V(\phi^{(0)}) = -{d (d-1) \ov L_{IR}^2}$, and  $V (\phi^{(0)}) < V(0)$,  $L > L_{IR}$. The flow solution then describes a domain wall with the metric~\eqref{1} interpolating between two AdS$_{d+1}$ with curvature radius given by $L$ and $L_{IR}$ respectively, i.e. 
\be \label{pep1}
f(z)  \to  {L^2 \ov L_{IR}^2} > 1 , \qquad z \to +\infty      \ .
\ee
 Near the IR fixed point, i.e. $z \to \infty$, $f$ can be expanded as 
 \be \label{find1}
 f (z) = {L^2 \ov L_{IR}^2} - {1 \ov (\tilde \mu z)^{2 \tilde \al}} + \cdots
 \ee
where $\tilde \al = \De_{\rm IR} - d$ with $\De_{\rm IR}$ the dimension of the leading irrelevant perturbing operator 
at the IR fixed point and $\tilde \mu$ is a mass scale characterizing irrelevant perturbations.  

\item Flow to a ``gapped'' phase: since $f$ has to increase with $z$, instead of approaching a constant as in~\eqref{pep1}, $f(z)$ can blow up as $z \to \infty$, e.g. 
\be \label{singf}
f(z) \propto z^{n}, \qquad n > 0 \ .
\ee
The spacetime then becomes singular at $z=\infty$. Given $n >0$, 
the $z= \infty$ singularity in fact sits at a {\it finite proper distance} away. 
From the standard IR/UV connection we expect that the system should be described by a gapped phase in the IR.
Explicit examples include the GPPZ~\cite{Girardello:1999} and Coulomb branch 
flow~\cite{Freedman:1999gk,Brandhuber:1999hb} which we will discuss in more detail in next section. 
While one should be wary of such singular spacetimes, they appear to give sensible answers for correlation functions consistent with the interpretation of a gapped phase (see e.g.~\cite{Bianchi:2001de}).\footnote{In Coulomb branch flow~\cite{Freedman:1999gk,Brandhuber:1999hb} which describes a Higgs phase of the $\sN=4$ SYM theory, there is a single Goldstone mode corresponding to spontaneous breaking of conformal symmetry. In the large $N$ limit, the effect of such a gapleess mode on observables like entanglement entropy can be neglected. We will thus still call it a ''gapped'' phase.}
In this paper we will assume such singular geometries make sense.

\een

\subsection{Holographic entanglement entropy}

\begin{figure}[h]
\begin{center}
\includegraphics[scale=0.5]{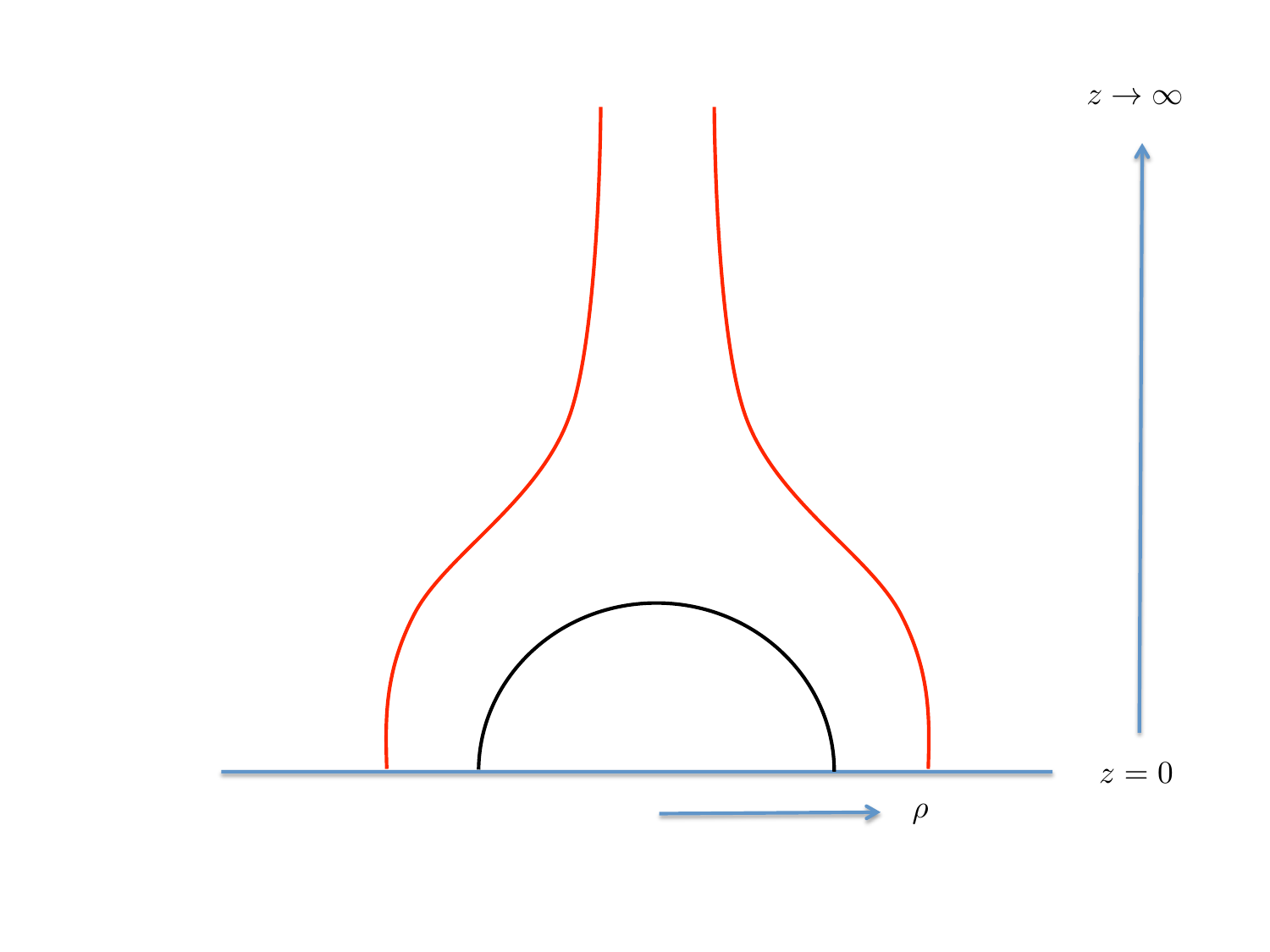} \quad
\end{center}
\caption{Cartoon of a minimal surface of disk topology (black) v.s. a minimal 
surface of cylinder topology~(red). The cylinder type surface is possible only for~\eqref{singf} with $n > 2$. 
}
\label{fig:cartoon}
\end{figure}

From the prescription of~\cite{Ryu:2006bv}, the entanglement entropy for a spherical region of radius $R$ is obtained by
\be 
S (R)= {2 \pi  L^{d-1} \ov \ka^2} \om_{d-2} A =  K A , \quad K \equiv {2 \pi  L^{d-1} \ov \ka^2} \om_{d-2}
\ee
where $\om_{d-2}$ is the area of a unit $(d-2)$-dimensional sphere and $A$ is obtained by minimizing  the surface area
\be 
A = \int_0^{z_m} d z \, {\rho^{d-2} \ov z^{d-1}} \sqrt{\rho'^2 + {1 \ov f (z)}} 
\label{minar} 
\ee
with the boundary condition at infinity
\be \label{imbd}
\rho (z=0) = R \ .
\ee
Depending on the spacetime metric, there can be two kinds of minimal surfaces as indicated in Fig.~\ref{fig:cartoon}. For the disk type, the minimal surface ends at a finite $z_m$ with 
\be \label{bd1}
\rho (z_m) = 0, \qquad \rho'(z_m) = \infty \ .
\ee
The cylinder type solution extends all the way to $z= \infty$ with 
\be \label{bd2}
\rho (z) \to \rho_0 , \quad \rho'(z) \to 0, \qquad z \to \infty
\ee
with $\rho_0$ a finite constant. When the IR geometry is given by another AdS$_{d+1}$ as in~\eqref{pep1}, the minimal surface is always of disk type. In Appendix~\ref{app:A} we show that 
for singular IR geometries with~\eqref{singf}, cylinder type solution can exist when the exponent $n>2$.   
As also discussed in Appendix~\ref{app:A}, for $n> 2$, there exists a critical minimal surface 
which closes off exactly at $z= \infty$, i.e. it can be viewed either as a surface of cylinder topology with $\rho_0 = 0$ or a surface of disk topology with $z_m = \infty$. This critical surface will be important for 
 ``second order phase transitions'' in entanglement entropy we observe in some examples of Sec.~\ref{sec:num}. 
 
At the UV fixed point, $f = 1$, the minimal surface is given by the hemisphere~\cite{Ryu:2006ef} 
\be \label{semi}
\rho_0 (z) =  \sqrt{R^2 - z^2}, \qquad z_{m0} = R \ .
\ee
Evaluating on~\eqref{semi}, the action~\eqref{minar} is divergent near $z=0$. We introduce a UV cutoff $z = \de$ and the regularized action becomes 
\be 
A = R \int_\de^R d z \, {(R^2 - z^2)^{d-3 \ov 2} \ov z^{d-1}}  \ .
\ee
One then finds the entanglement 
entropy has the form~\eqref{enD}, with 
\be \label{cftR} 
\sS_d^{\rm (UV)} = s_d^{\rm (UV)} =  
{(d-3)!! \ov (d-2)!!} K  \ .
\ee
Similarly at the IR fixed point with $f = {L^2 \ov L_{\rm IR}^2}$  we find 
\be \label{cftir}
\sS_d^{\rm (IR)} = s_d^{\rm (IR)} = {(d-3)!! \ov (d-2)!!} K_{\rm IR} \ , \quad K_{\rm IR} \equiv  {2 \pi  L_{\rm IR}^{d-1} \ov \ka^2} \om_{d-2} \ .
\ee

\subsection{$\sS_d$ in terms of asymptotic data} \label{sec:asym}

We now derive a formula to relate
$\sS_d$ directly to a finite coefficient in the asymptotic expansion of the minimal surface solution near the boundary. For definiteness we will restrict our discussion to a single scalar field, but the conclusion is  general. 

Consider the action~\eqref{minar} with a cutoff at $z= \de$,
\be \label{cuact}
A = \int_\de^{z_m}  dz \, {\rho^{d-2} \ov z^{d-1}} \sqrt{\rho'^2 + {1 \ov f}}
\equiv \int_\de^{z_m} dz \,  \sL  \ 
\ee
with boundary conditions~\eqref{imbd} and~\eqref{bd1} (or~\eqref{bd2} with $z_m = \infty$). In the above equation, varying $R$ with $\de$ fixed, and using the standard Hamilton-Jacobi method we find 
\be \label{erp}
{d  A \ov d R}=- \sH (z_m) {d z_m \ov dR} - \Pi {d \rho \ov dR} \biggr|_{\de}
= - \Pi (\de) {d \rho (\de) \ov dR}
\ee
where 
\be \label{canoD}
\Pi = {\p \sL \ov \p \rho'} =  {\rho^{d-2} \ov z^{d-1}} {\rho' \ov \sqrt{\rho'^2 + {1 \ov f}}}, \qquad
\sH = \Pi \rho' - \sL = - {\rho^{d-2} \ov z^{d-1}} {1 \ov f \sqrt{\rho'^2 + {1 \ov f}}}
\ee
and we have used that $\sH (z_m) =0$ due to the boundary condition there (for both~\eqref{bd1} and~\eqref{bd2}). Equation~\eqref{erp} only depends on the local solution near infinity and can be evaluated using the asymptotic expansion 
there. 

For small $z$, $f(z)$ has the expansion (see e.g. also~\cite{Hung:2011ta}) 
\be \label{fexp1}
f(z) = 1+ \sum_{n=0}^\infty f_n z^{d+n} + \sum_{m=2}^\infty c_n z^{m \al}
\ee 
where $\al$ was introduced in~\eqref{find}. Note that here we are considering a flat boundary, so the first series of~\eqref{fexp1} starts only at order $O(z^d)$. The second series comes from the scalar contribution. Similarly, $\rho(z)$ can be expanded as~\cite{Graham:1999pm,Schwimmer:2008yh,Hung:2011ta}\footnote{There are also logarithmic terms when two series share common terms. For example, when $n + m \al = d$, there is a term proportional to $z^d \log \mu z$  with $\mu$ a mass scale associated with relevant perturbation. Presence of these additional logs will not affect our discussion. So below we will not make them explicit.}
\be  \label{imexp}
\rho(z) = R  - {z^2 \ov 2R} + \cdots + c_d (R) z^d + \cdots + \sum_{n=2,m=2}^\infty a_{nm} (R) z^{n+m \al} \ .
\ee
In~\eqref{imexp}, the first $\cdots$ contains  only  even powers of $z$ up to $z^d$ and
the second  $\cdots$ contains integer powers greater than $d$. Note that the structure of expansion in~\eqref{imexp} excluding the last term follows from the standard Fefferman-Graham expansion applied to a submanifold in AdS~\cite{Graham:1999pm}, while the last term comes from the similar series in the expansion of $f$.  In~\eqref{imexp},  $c_d$ is undetermined and all other coefficients can be determined in terms of coefficients in~\eqref{fexp1} and $c_d$ from local analysis of the minimal surface equation of motion near $z=0$. 

Plugging the expansion~\eqref{imexp} into~\eqref{erp} we find that  in the limit $\de \to 0$
\be \label{univE}
{dA \ov dR} =  - d R^{d-2} c_d (R) - {e_d \ov R} + {d A \ov dR} \biggr|_{\rm non-universal}
 \ee
where 
\be
e_d = \ha \le(1 + (-1)^d \ri) \le(1 - (-1)^{d-2 \ov 2} \ri) {(d-3)!! \ov (d-2)!!} 
 \ee
 in non-vanishing only in $d=4k, k=1,2,\cdots$ dimensions.  The non-universal part has the form 
 \be \label{nonu}
 {d A \ov dR} \biggr|_{\rm non-universal} = b_1 (\de) R^{d-3} + b_2 (\de) R^{d-5} + \cdots 
 \ee
with only terms of non-negative powers of $R$. $b_n$ can be expressed in terms of a divergent series of $\de$, with $\de^{2-d}$ the most divergent term in $b_1$, $\de^{4-d}$ for $b_2$, and so on.   More explicitly,  introducing a dimensional parameter $\mu$ to characterize the dimension of $c_n$ (with $c_n \sim \mu^{n \al}$), then the series in $b_1$ can be schematically written as 
\be \label{bexp}
b_1 \sim {1 \ov \de^{d-2}} \le(1 + (\mu \de)^{2 \al} + (\mu \de)^{3 \al} + \cdots \ri) + {\rm const} \ .
\ee
The structure of non-universal terms~\eqref{nonu} are consistent with that 
argued in Sec.~\ref{app:UV} (recall the discussion below~\eqref{yuep}).
They go away when acted with the differential operator in~\eqref{Scen} and do not appear in $\sS_d$.  

As a result $\sS_d (R)$ can be solely expressed in terms of $c_d (R)$.
 More explicitly, from~\eqref{univE} and~\eqref{c3} we find that for $d=3$
\be \label{rr1}
{1 \ov K} \sS_3 (R)= - 3 R^2 c_3 (R) + 3 \int_0^R dR' \, R' c_3 (R')  + C 
\ee
where $C$ is determined by requiring $\sS_3 (R=0)$ reduces to the value at the UV fixed point. 
Also note 
\be 
{1 \ov K} {d \sS_3 \ov dR} = - 3 R\, \p_R \le[R c_3 (R)\ri] \ .
\ee
Similarly, for $d=4$, 
\be \label{rr2}
{1 \ov  K} \sS_4 = 1  -  2 R^3 c_4 (R) -   2 R^4 {d c_4 \ov dR}  \ . 
\ee

\subsection{Two closely separated fixed points}  \label{sec:pof1}

Let us now consider the situation in which the bulk cosmological constants of the UV and IR fixed points 
are close to each other, which translates into the boundary theory as that the central charges $s_d$~\eqref{cftR} of the UV and IR fixed points are close. In this case we have
\be \label{neke}
f(z) = 1 + \ep g (z), \qquad \ep \equiv {L^2 \ov L_{\rm IR}^2} - 1 \ll 1, \qquad g' (z) > 0, 
\qquad g (z) \to  \bca  0 & z \to 0 \cr
                                    1 & z \to \infty
                                    \eca  \ .
\ee
We expect the minimal surface to be close to that of a CFT, i.e. 
\be 
\rho(z) = \rho_0 (z) + \ep \rho_1(z)+\cdots 
\ee
with $\rho_0$ given by~\eqref{semi}, and 
\be \label{aexp}
A = A_0 + \ep A_1 + \cdots   \ .
\ee 

Under variations of $f$, the variation of the cutoff action~\eqref{cuact} can be written as  
\be 
\de A = \int_{\de}^{z_{m}} d z \,  {\de \sL \ov \de f} \de f - \sH (z_m) \de z_m 
 - \Pi \de \rho  \bigr|_{\de} 
\ee
where $\sH$ and $\Pi$ were introduced in~\eqref{canoD} and $\de z_m$ and $\de \rho$ denote the induced variations due to $\de f$. Now apply the above equation to~\eqref{neke}--\eqref{aexp}, we find that 
\be \label{w0}
A_1 = \int_{\de}^{R} dz \,  {\de \sL \ov \de f}\biggr|_{\rho_0}  \, g (z)  + \Pi (\de) \bigr|_{\rho_0} \rho_1 (\de)
\ee
where $\sH(z_m)$ again vanishes due to the boundary condition~\eqref{imbd} at $z_m$. $\rho_1 (z)$ is obtained by solving the linearized equation following from~\eqref{minar} around~\eqref{semi}. To find its value at the short-distance cutoff $\de$, it is enough to use the leading order expression~\eqref{find} for $f$ (with $g(z) = (\mu z)^{2 \al} + \cdots$), and we find that\footnote{When $\al = {d-2 \ov 2}$, equation~\eqref{exz1}  becomes $\rho_1 (\de)= - {d-2 \ov 2d} (\mu \de)^{d-2} {\de^2 \ov \sqrt{R^2-\de^2}} \log \mu \de$.}
\be \label{exz1}
\rho_1 (\de) =c (\al) (\mu \de)^{2\al}  {\de^2 \ov \sqrt{R^2 - \de^2}} + \cdots, \qquad c(\al) = {2+ \al - d \ov 2 (1 + \al) (2 + 2 \al - d)} \ .
\ee

Plugging the explicit form~\eqref{semi} of $\rho_0$ and~\eqref{exz1} into~\eqref{w0}, we find that 
\be \label{eoop}
A_1= - {1 \ov 2R}  \int_{\de}^R d z  \, (R^2 -z^2)^{d-1 \ov 2} \, z^{1-d}  \, g(z)
+  c (\al)  {(R^2 - \de^2)^{d-3 \ov 2} \ov R \de^{d-4}} (\mu \de)^{2\al} + \cdots  \ .
\ee 
The divergent terms in the above equation are again consistent with the structure described in Sec.~\ref{sec:div} and thus will not contribute to $\sS_d (R)$.\footnote{Note that for some special values of $\al$, both terms in the above equation can lead to logarithmic terms which contain $\log \mu \de$, as pointed out before in~\cite{Hertzberg:2010uv,Hung:2011ta}. These terms are also consistent with the structure described in Sec.~\ref{app:UV}.} 

Now applying~\eqref{Scen} to~\eqref{eoop} we find that, for odd $d$, 
\be \label{master1}
\sS_{d}^{(1)} (R)= 
   -  {(d-1)!! \ov (d-2)!!}  {K \ov 2R} \int_{0}^R dz \, g(z)  = -  {(d-1)!! \ov (d-2)!!}  {K \ov 2} \int_{0}^1 dx \, g(xR)
\ee
and for even $d$
\be \label{master2}
 \sS^{(1)}_d  (R)= -  {(d-1)!! \ov (d-2)!!}  {K \ov 2R} \int_{0}^R dz \, {z g(z) \ov \sqrt{R^2 -z^2}}  
 = -  {(d-1)!! \ov (d-2)!!}  {K \ov 2} \int_{0}^1 dx \, {x g(xR) \ov \sqrt{1 -x^2}}  \ .
 \ee
 It is interesting to note that for a monotonic $g(z)$, both~\eqref{master1} and~\eqref{master2} are
 monotonic
\be 
{d \sS^{(1)}_d \ov d R} < 0 \ .
\ee
Also note that in the limit $R \to \infty$, we can replace $g(xR)$ in both expressions by its value 
at infinity $g (\infty) = 1$, leading to  
\be  \label{firpr}
\sS^{(1)}_d  (R \to \infty) =  -  {(d-1)!! \ov (d-2)!!}  {K \ov 2} 
\ee
which is precisely the expected difference between the values of the IR and UV fixed point (see~\eqref{cftR} and~\eqref{cftir}) expanded to first order in $\ep$. Thus to first order in $\ep$ we have recovered 
\be  \label{lagr0} 
\sS_d (R \to \infty) = \sS_{d}^{\rm (IR)}  \ .
\ee


\subsection{Leading small and large $R$-dependence} \label{sec:Rasy}

In this section we examine the leading small and large $R$-dependence of $\sS_d (R)$. 

When $R$ is small, the range 
of $z$ covered by the minimal surface is also small. Thus we can use the asymptotic form~\eqref{find} of the function $f(z) = 1 + \de f + \cdots $ with $\de f = (\mu z)^{2 \al}$, and treat the deviation from pure AdS as a small perturbation. To first order in $\de f$, equation~\eqref{w0} then again applies with $g$ there replaced by $\de f$. The calculation then becomes very similar to that of the last subsection and we find the following leading behavior 
\be \label{smlRB}
\sS_d (R) = \sS_d^{(\rm UV)} -  {(d-1)!! \ov (d-2)!!}  {K \ov 2}  a (\al)  (\mu R)^{2 \al} + \cdots ,\quad R \to 0
\ee
with $\sS_d^{(\rm UV)}$ given by~\eqref{cftR} and $a (\al)$ is given by
\be \label{aval}
a (\al) = \bca 
      {1 \ov 1 + 2 \al} & {\rm d \; odd} \cr
      { \sqrt{\pi} \Ga (1+\al)\ov 2 \Ga \le({3 \ov 2} + \al \ri)} 
      & {\rm d \; even} 
      \eca  \ .
      \ee
Recall that for a source deformation $\al = d -\De$ where $\De$ is the UV dimension of the leading relevant operator. Equation~\eqref{smlRB} has a simple interpretation that the leading contribution from a relevant operator comes at two-point level, and should be valid also outside holographic systems.
Also note 
that the sign of the numerical coefficient in~\eqref{smlRB} is such that at this order $\sS_d (R)$ is monotonically decreasing with $R$.

In the large $R$ limit,  estimating the leading $R$-dependent correction to the asymptotic value $\sS_d (R \to \infty) = \sS_{d}^{\rm (IR)}$ for a general holographic system becomes more challenging. 
Here we will consider the case of two closely separated fixed points discussed in the last subsection  where we can take advantage of the closed expressions~\eqref{master1} and~\eqref{master2}.
  At large $R$, the argument $xR$ for $g(xR)$ is large except near $x=0$,  thus we should be able to use the asymptotic expansion of $g(z)$ for large $z$ (from~\eqref{find1} and~\eqref{neke}), 
  \be \label{gexp}
  g(z) = 1 - (\tilde \mu z)^{- 2 \tilde \al} + \cdots \ .
  \ee
 Recall that $\tilde \al = \De_{\rm IR} - d$ where $\De_{\rm IR}$ is the IR scaling dimension of the leading irrelevant operator at the IR fixed point. Plugging~\eqref{gexp} into ~\eqref{master1} and~\eqref{master2}, the $1$ in~\eqref{gexp} will give~\eqref{firpr}. To to first order in $\ep$ we can write 
 \be 
\sS_d (R) = \sS_{d}^{\rm (IR)} + I_d (R) + O(\ep^2) 
 \ee
 where 
 \be \label{iexp1}
 I_d (R) =   {(d-1)!! \ov (d-2)!!}  {\ep K \ov 2} \int_{0}^1 dx \,  {1 \ov (\tilde \mu R x)^{2 \tilde \al}} 
 \bca 
     1  + \cdots& {\rm odd \; d}  \cr
      {x  \ov \sqrt{1 -x^2}} + \cdots  & {\rm even \; d}
      \eca    \ .
 \ee
 In~\eqref{iexp1} the $\cdots$ denotes contributions from higher order terms in~\eqref{gexp} (represented by $\cdots$ there). Now notice that the integral in~\eqref{iexp1} becomes divergent when $\tilde \al \geq \ha$  for odd $d$ and $\tilde \al \geq 1$ for even $d$. In fact, even if the leading term in~\eqref{iexp1} is convergent, higher order terms in~\eqref{gexp} will still eventually lead to 
 divergent integrals (and will be more and more divergent). Physically such divergences reflect that  higher order terms in the expansion of $g$ are more irrelevant and thus are more sensitive to the short-distance structure near $x=0$. 

When~\eqref{iexp1} is convergent we find 
 \be \label{LmR}
 I_d (R) = {(d-1)!! \ov  (d-2)!!}  { \ep K \ov 2}  {b (\tilde \al) \ov (\tilde \mu R)^{2 \tilde \al} }+ \cdots 
 \ee
with
\be \label{valb}
b  (\tilde \al)= \bca  {1 \ov 1 - 2 \tilde \al} & \tilde \al < \ha, \quad {\rm d \; odd} \cr
     {\sqrt{\pi} \Ga (1-\tilde \al) \ov 2 \Ga ({3 \ov 2} - \tilde \al)}  &  \tilde \al < 1, \quad {\rm d \; even}
\eca \ .
\ee


Now using odd $d$ for illustration, consider $\tilde \al > \ha$.  Since the large $R$ expansion of 
$g$ assumes that $\tilde \mu z \gg 1$, we should split the integral at $x  \sim  {1 \ov \tilde \mu R}$, i.e. 
\be \label{spt}
\int_0^1 dx \, g(xR) = \int_0^{1 \ov \tilde \mu R} g (xR) + \int_{1 \ov \tilde \mu R}^1 dx g (xR) \ .
\ee
Since $g(z)$ is well defined near $z=0$, the first term in the above is convergent and is of order $O( {1 \ov \tilde \mu R})$. Now plugging the large $R$ expansion of $g$ into the second term of~\eqref{spt} we find the previously divergent term becomes 
\be
{1 \ov (\tilde \mu R)^{2 \tilde \al}} \int_{1 \ov \tilde \mu R}^1 dx  \, x^{- 2 \tilde \al} \sim {1 \ov \tilde \mu R} + \cdots\ .
\ee
Similarly, the leading contribution of a term of the form $z^{-n}$ with $n > 1$ in the 
large $z$ expansion of $g(z)$  will always be of order $O(R^{-1})$ regardless of the value of $n$. 
We thus conclude that for $\tilde \al > \ha$
\be 
I_{d} (R, \tilde \al) \sim {1 \ov \tilde \mu R} 
\ee
where the numerical coefficient now  receives the contribution from all orders of the large $z$ expansion of $g(z)$, and the validity of the above expression assumes that this infinite sum is convergent. When $\tilde \al < \ha$, then the leading dependence is given by~\eqref{LmR} as  potentially divergent higher order terms give only a contribution which is of order $O({1 \ov \tilde \mu R})$ and is subleading compared to~\eqref{LmR}. 

Exactly the same can be said regarding~\eqref{master2} for even $d$ except that the range becomes $\tilde \al > 1$ and the leading order contribution  becomes of order $O(R^{-2})$. 

We thus conclude that
\be \label{lmr0}
  \sS_d  (R) =  \sS_{d}^{\rm (IR)}  + {(d-1)!! \ov  (d-2)!!}  { \ep K \ov 2} 
  {b (\tilde \al) \ov (\tilde \mu R)^{2 \tilde \al} } + O(\ep^2),
  \quad \tilde \al < \bca \ha &  {\rm d \; odd} \cr
              1 & {\rm d \; even} \eca 
                \ee
with $b (\tilde \al)$ given by~\eqref{valb}, and  
\be \label{lmr1}
  \sS_d  (R) =  \sS_{d}^{\rm (IR)}  + {(d-1)!! \ov  (d-2)!!}  { \ep K \ov 2} \bca 
   O \le({1 \ov \tilde \mu R} \ri)  & \tilde \al > \ha \;({\rm odd \; d}) \cr
   O \le({1 \ov (\tilde \mu R)^2} \ri)  & \tilde \al > 1 \;({\rm even \; d})  
   \eca  + O(\ep^2)\ .
  \ee
Note that equations~\eqref{lmr1} agree with the general expectation~\eqref{larr} of Sec.~\ref{app:UV}. 
Here we find that when the IR dimension of the involved irrelevant operator is not too large one should instead have~\eqref{lmr0}, which can be considered as predictions from the holographic duality.

Finally when $\tilde \al = \ha$ for odd $d$ and $\tilde \al =1$ for even $d$, similar argument leads to 
\be 
 \sS_d  (R) =  \sS_{d}^{\rm (IR)}  + {(d-1)!! \ov  (d-2)!!}  { \ep K \ov 2} 
   \bca  {\log \tilde \mu R \ov \tilde \mu R}   &  \tilde \al = \ha \; ({\rm odd \; d}) \cr
      {\log \tilde \mu R \ov (\tilde \mu R)^2}  &  \tilde \al = 1 \; ({\rm even \; d})
 \eca + \cdots \ .
 \ee


\section{Some numerical studies} \label{sec:num}

We now consider some specific  holographic RG flows, including some ``realistic'' ones which describe known field theory flows at large $N$ and strong coupling. In these examples, the minimal surface cannot be found analytically and numerical calculations are needed. In some examples, the gravity solutions are only known numerically.  In computing the entanglement entropy, one could to choose to evaluate~\eqref{minar} directly after obtaining the numerical solution for the minimal surface. The numerical integrations can sometimes be time-consuming. It is often more convenient to use~\eqref{rr1} or~\eqref{rr2} to obtain $\sS_d$ directly from the coefficient $c_d (R)$ in the asymptotic expansion of the minimal surface solution near the boundary.   

We will first consider $d=3$, where we find that $\sS_3 (R)$ is always non-negative and monotonic. This is no longer the case for $d=4$. The non-monotonic examples include the GPPZ flow~\cite{Girardello:1999} which describes the flow of the $\sN=4$ SYM theory to a  confining phase under a mass deformation and the so-called Coulomb branch flow~\cite{Freedman:1999gk,Brandhuber:1999hb} which describes the $\sN=4$ SYM theory in a Higgs phase.

The two realistic examples below in Sec.~\ref{sec:real1} and~\ref{sec:real2}  have also been considered recently in~\cite{Albash:2011nq}.

\subsection{$d=3$} 

\subsubsection{A  realistic flow} \label{sec:real1}

\begin{figure}[h]
\begin{center}
\includegraphics[scale=0.56]{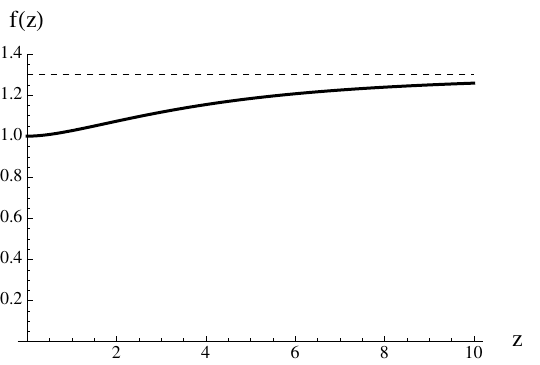} 
\includegraphics[scale=0.56]{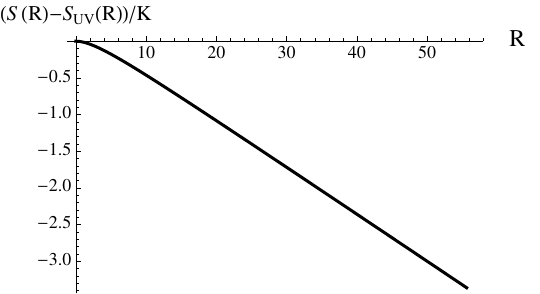} 
\includegraphics[scale=0.56]{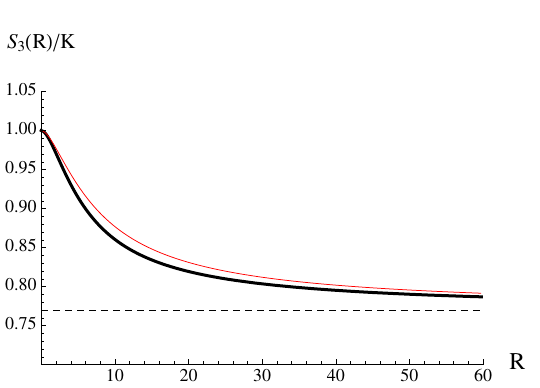}
\end{center}
\caption{{\bf Left:} $f(z)$ for the domain wall solution describing the flow of M2-brane theory to
an IR fixed point preserving $\sN=2$ supersymmetry. {\bf Middle:} plot of $S(R) - S_{\rm UV} (R)$ where $S_{\rm UV}$ denotes that at the UV fixed point. 
The UV divergences cancel when taking the difference, but the resulting expression does not have a well-defined large $R$ limit, with a linear $R$-dependence. As in the case of a free massive scalar and Dirac field of Sec.~\ref{sec:free}, the presence of such a linear term can be understood as a finite renormalization between the short distance cutoffs of the UV and IR fixed points, as discussed in Sec.~\ref{app:UV}.
{\bf Right:}  $\sS_3 (R)$ for this flow. We normalize the value at UV to be $1$. The horizontal dashed line denotes the expected value~\eqref{cftir} for the IR fixed point. The black line (lower line) is obtained from  numerical calculation by using~\eqref{rr1}. For this flow, $\ep$ in~\eqref{neke} is $\ep \approx 0.3$, and thus equation~\eqref{master1} should be a reasonable approximation, whose results are plotted using the red line (upper line). Note that the part linear in $R$ in $-S_{\rm finite}$ as seen in the second plot is automatically eliminated when considering $\sS_3 (R)$. 
}
\label{f3plot}
\end{figure}

We first consider the holographic RG flow of~\cite{Corrado:2001nv} which describes the flow 
from M2 brane theory at UV to an $\sN =2$ superconformal theory with an $SU(3) \times U(1)$ 
global symmetry at IR. The flow involves two scalar operators whose UV and IR dimensions are respectively 
\be 
\De = (2,2), \qquad  \De_{\rm IR} = \le(\ha (1+\sqrt{17}), \ha (5 + \sqrt{17}) \ri) \ .
\ee
Thus in~\eqref{find} and~\eqref{find1}, $\al = 1$ and $\tilde \al = \ha (\sqrt{17}-1) \approx 1.562$. 
Also ${L^2 \ov L_{UV}^2} = {3^{3 \ov 2} \ov 4} \approx 1.299$. 
The numerical results are presented in Fig.~\ref{f3plot}. Clearly, $\sS_3 (R)$ is positive, and monotonic. 
Note that for this flow, $\ep$ in~\eqref{neke} is $\ep \approx 0.3$, and thus equation~\eqref{master1} should be a reasonable approximation, which is confirmed by numerical results.
Since $\tilde \al > \ha$, from~\eqref{lmr1} we expect $\sS_3 (R)$ should fall off as $1/R$ for large $R$. This appears to fit the numerical data very well.

\subsubsection{A sharp domain wall: first order phase transition} \label{sec:shp3d}

\begin{figure}[h]
\begin{center}
\includegraphics[scale=0.56]{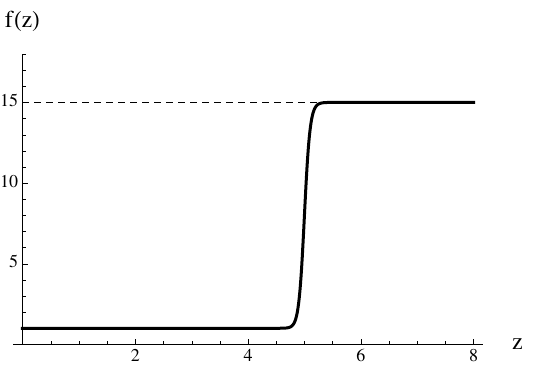} 
\includegraphics[scale=0.6]{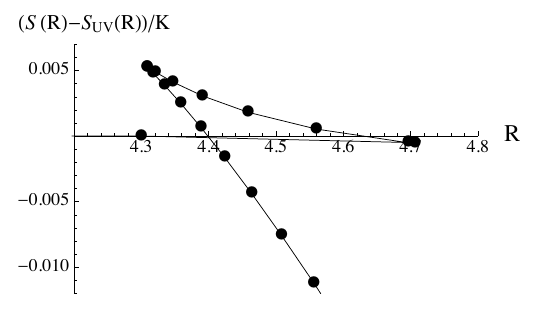} 
\includegraphics[scale=0.6]{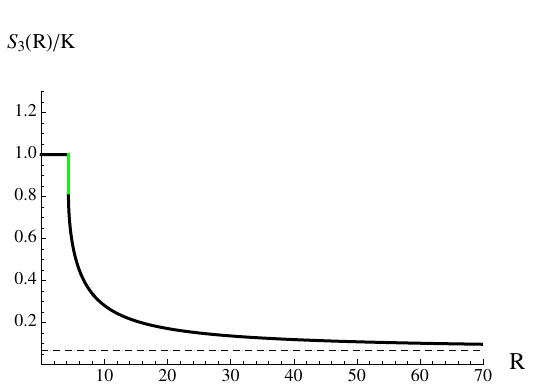}
\end{center}
\caption{{\bf Left:} a steep domain wall (toy example) with $f (z) = 1 +  {14 z^{100}  \ov 5^{100} + z^{100}}$. {\bf Middle:} plot of $S (R)- S_{\rm conf} (R)$ where $S_{\rm conf} (R)$ denotes the entanglement entropy for the UV fixed point. The short-distance divergences cancel when taking the differences.  For the indicated range of $R$, the action~\eqref{minar} has three extrema, all of disk type. The entanglement entropy of the system is given by the smallest of them. There is a first-order ``phase transition'' at $R_c=4.4$. 
{\bf Right:}  $\sS_3 (R)$ has a discontinuous jump, which is indicated by the vertical green line. The dashed horizontal line is the expected asymptotic value for the IR fixed point. }
\label{tplot1}
\end{figure}

The earlier example is a rather shallow domain wall. Let us now consider a sharp domain wall. Unfortunately, there appears no realistic example of this type. We will thus play with toy examples by coming up with various monotonic functions $f (z)$, a strategy which we will repeatedly use below.  
By scanning through various examples we find that even for ${L \ov L_{\rm IR}} \gg 1$, $\sS_3 (R)$ remains monotonic and approaches to the right asymptotic value. However, a new phenomenon arises when the domain wall becomes steep, as in the left plot of Fig.~\ref{tplot1}. In this case, 
the minimal surface action~\eqref{minar} can have more than one extrema for certain range of $R$, as indicated in the middle plot of Fig.~\ref{tplot1}.  The minimal surface prescription instructs us to pick the one with the smallest area. While the entanglement entropy itself is continuous as a function of $R$, after taking derivative to obtain $\sS_3 (R)$, we find there is a discontinuous jump at some value $R_c$  as shown in the right plot of Fig.~\ref{tplot1}. In other words, there  is a first-order ``phase transition'' in the entanglement entropy at $R_c$. Despite the jump, $\sS_3$ is still monotonic and approaches the right asymptotic value.

\subsubsection{``Gapped'' phases} \label{sec:fgap}

\begin{figure}[h]
\begin{center}
\includegraphics[scale=0.56]{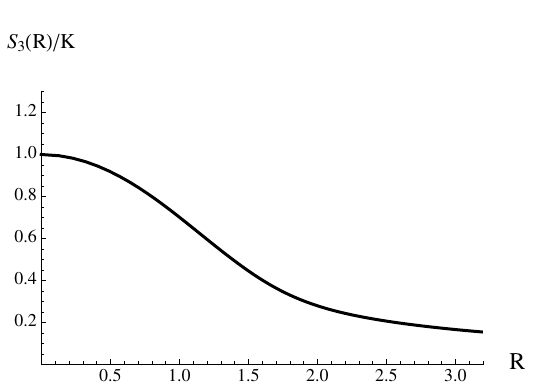} \quad
\includegraphics[scale=0.56]{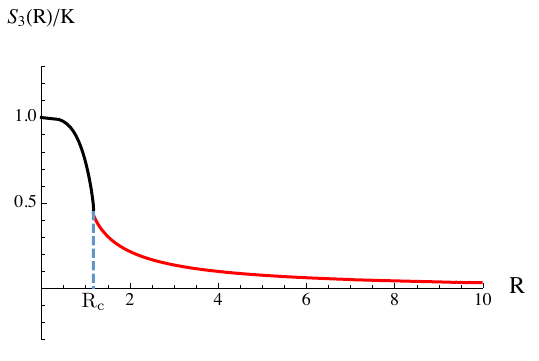} \quad
\includegraphics[scale=0.56]{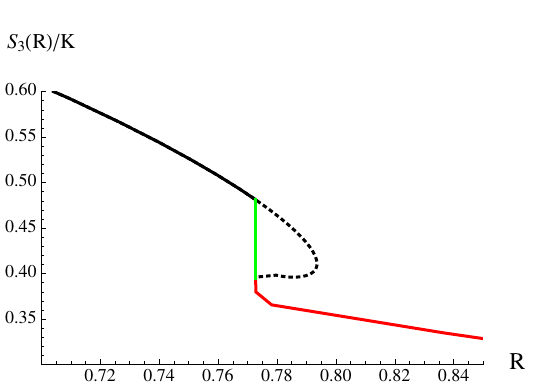}
\end{center}
\caption{{\bf Left:} $\sS_3 (R)$ for $f(z) = 1 + z^2$.  {\bf Middle:} $\sS_3 (R)$ for $f (z) = 1 + z^3$ which exhibits a ``second-order phase transition'' from minimal surface of disk topology (black curve) to cylindrical topology (red curve). {\bf Right:}  $\sS_3 (R)$ for $f (z) = (1 + z^2)^2 $, which exhibits a ``first-order phase transition'' between the surfaces of two topologies. The dashed curve corresponds to other extrema of the minimal surface action. There is a discontinuous jump in $\sS_3 (R)$ which is indicated by the green vertical line.}
\label{tplot2}
\end{figure}

We now consider some examples where the IR is described by a ``gapped'' phase. We are interested in systems with Lorentz symmetry, and thus are forced to consider singular geometries discussed around~\eqref{singf}.  An example of  gapped phase with a regular geometry is the AdS soliton~\cite{Witten:1998zw}. But the corresponding boundary theory contains a circle direction. Our discussion of the divergence structure of the entanglement entropy in Sec.~\ref{app:UV} thus does not apply, and one needs to modify the definition of $\sS_d (R)$, which we will not pursue here. 
 The holographic entanglement entropy for an AdS soliton was considered before in~\cite{Pakman:2008ui}. 
A first-order phase transition between minimal surfaces with disk and cylinder topology similar to our third example below (right plot of Fig.~\ref{tplot2}) was found there.

 Again due to lack of simple explicit examples, we will consider some toy examples by postulating certain $f(z)$. 
In Fig.~\ref{tplot2}, we consider three such examples which illustrate three possible scenarios. 

In the left plot we consider $f(z) = 1 + z^2$. In this case since $f(z \to \infty) \sim z^2$ with an exponent $2$, only minimal surface of disk topology is possible. There appears to be a unique minimal surface solution for all $R$, and $\sS_3 (R)$ is smooth. 

The middle plot is for $f(z) = 1 + z^3$, in which case, when $R$ is sufficiently large, the minimal surface is of cylinder type (see discussion around~\eqref{bd2} and Appendix~\ref{app:A}). This is indicated in the plot by the red curve. 
The black curve for smaller $R$ values has minimal surface of disk type. $\sS_3 (R)$ appears to be continuous at the transition point where the minimal surface changes topology. Thus the entanglement entropy has a ``second-order phase transition.'' Note that approaching the transition point from the smaller $R$ side, the end point $z_m$ of the disk-like surface approaches infinity, while from the larger $R$ side, $\rho_0$ for the cylinder-like surface approaches zero. The two branches meet at the critical 
surface discussed in Appendix~\ref{app:A} (see Fig.~\ref{fig:cartoon2} for a cartoon).
It may be interesting to understand the critical behavior of such a transition which we will leave for future study.

The right plot of Fig.~\ref{tplot2} is for $f(z) = (1 + z^2)^2$, which exhibits a ``first-order phase transition.'' Again the red and black curves have minimal surface of cylinder-type and disk-type respectively. Now near the transition between the two topologies, the action~\eqref{minar} now has three extrema, with the non-minimal area extrema indicated in the plot by dotted lines. There is a discontinuous jump in $\sS_3 (R)$ at the transition point. 

In all these examples, $\sS_3 (R)$ appears to approach zero at large $R$, although our numerics cannot go to too large $R$.

\begin{figure}[h]
\begin{center}
\includegraphics[scale=0.5]{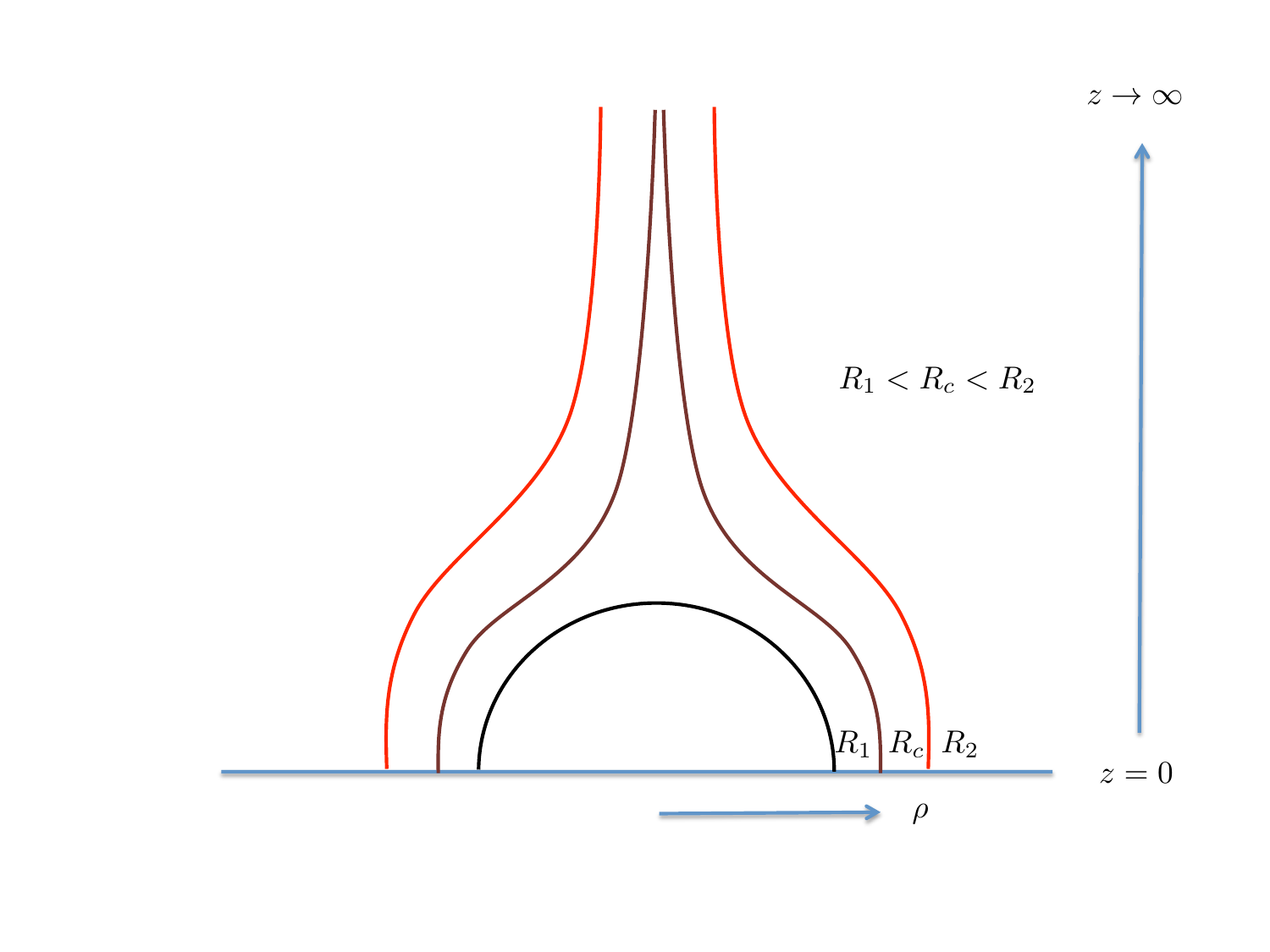} \quad
\end{center}
\caption{Cartoon of a ``second-order phase transition'' from a minimal surface of disk topology (black) to a minimal 
surface of cylinder topology~(red) as we increase $R$. The surface (brown) at $R_c$ is the critical surface discussed in Appendix~\ref{app:A}.
}
\label{fig:cartoon2}
\end{figure}

\subsubsection{Non-monotonic $f(z)$} 

\begin{figure}[h]
\begin{center}
\includegraphics[scale=0.56]{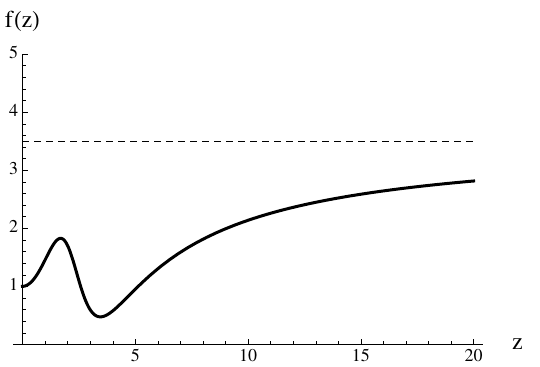} 
\includegraphics[scale=0.56]{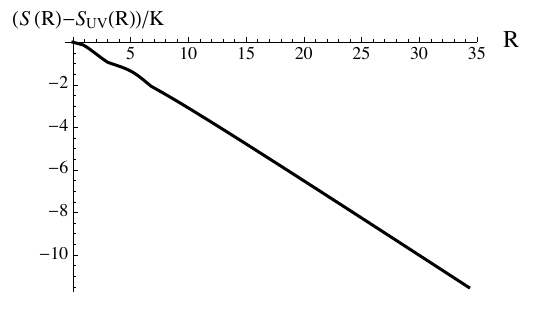}
\includegraphics[scale=0.56]{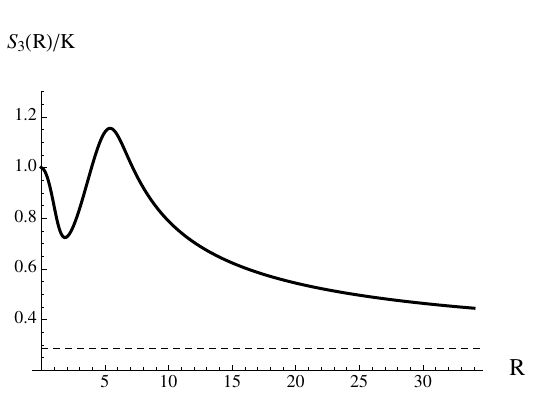}
\end{center}
\caption{{\bf Left:} A toy model of $f(z)$ which is non-monotonic. {\bf Middle:} The entanglement entropy itself is not sensitive to the non-monotonic feature of $f$. {\bf Right:} the corresponding  $\sS_3 (R)$ is sensitive and also becomes non-monotonic.}
\label{toy}
\end{figure}

Finally just for comparison, in Fig.~\ref{toy} we show a toy model of $f$ which is not monotonic, neither is the corresponding $\sS_3$. This toy example illustrates nicely several important aspects of the properties of $\sS_3 (R)$:

\ben 

\item  As mentioned earlier, the monotonicity of $f$ is tied to the imposing of null energy condition, which can be interpreted as the bulk reflection of boundary unitarity.  
This indicates that the monotonicity of $\sS_3 (R)$ is closely tied to unitarity of the boundary system. 

\item The location in $R$ where the non-monotonicity of $\sS_3 (R)$ occurs appears to roughly scale with $z$. From the IR/UV connection, $z$ translats into a boundary length scale. That the non-monotonicity of $f(z)$ at certain $z$ directly translates into non-monotonicity of $\sS_3 (R)$ at certain  $R$ which roughly scales with $z$ indicates that $\sS_3 (R)$ is most sensitive to contributions from degrees of freedom around scale $R$. In contrast, as shown in the figure the entanglement entropy itself is rather featureless 
as a function of $R$. 

\een

\subsection{$d=4$}

\subsubsection{Leigh-Strassler flow} \label{sec:real2}

\begin{figure}[h]
\begin{center}
\includegraphics[scale=0.8]{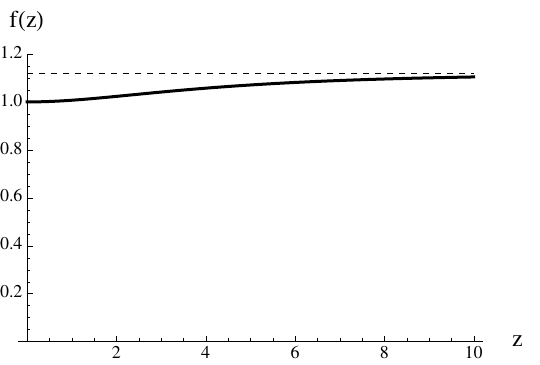} \qquad
\includegraphics[scale=0.8]{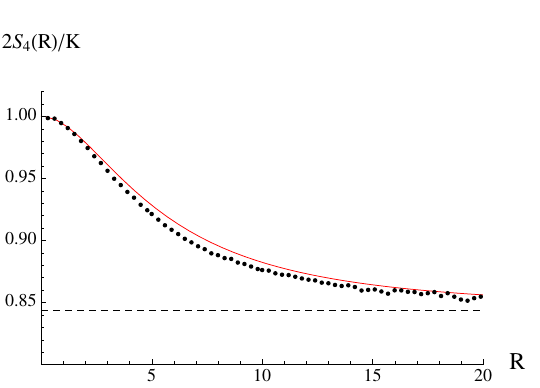}
\end{center}
\caption{{\bf Left:} $f(z)$ for the domain wall solution describing the Leigh-Strassler flow from $\sN=4$ SYM theory to an IR fixed point preserving $\sN=1$ supersymmetry. {\bf Right:}  $\sS_4 (R)$ for this flow. The horizontal line denotes the expected value for the IR theory. The dotted line is obtained from direct numerical calculation. For this flow, $\ep$ in~\eqref{neke} is $\ep \approx 0.12$, and thus equation~\eqref{master2} should be a good approximation, whose results are plotted using the solid line.  }
\label{f4plot}
\end{figure}

We first consider the flow from the $\sN=4$ SYM theory ($d=4$) to an $\sN=1$ superconformal fixed point with $SU(2) \times U(1)$ global symmetry~\cite{Khavaev:1998fb,Freedman:1999gp}. The flow involves two scalar operators whose UV and IR dimensions are respectively 
\be 
\De = (2,3), \qquad \De_{\rm IR} = \le(1+\sqrt{7}, 3+\sqrt{7} \ri) \ .
\ee
Thus in~\eqref{find} and~\eqref{find1}, $\al = 1$ and $\tilde \al = \sqrt{7}-1 \approx 1.646$. 
Also ${L^2 \ov L_{UV}^2} = {2^{10 \ov 3} \ov 9} \approx 1.12 $. 
The numerical results are given in Fig.~\ref{f4plot}.  $\sS_4 (R)$ is  
positive and monotonic. 
For this flow, $\ep$ in~\eqref{neke} is $\ep \approx 0.12$, and thus equation~\eqref{master2} should be a good approximation, which is confirmed by the numerical results.
Since $\tilde \al > 1$, from~\eqref{lmr1} we expect $\sS_4 (R)$ should fall off as $1/R^2$ for large $R$.  Our numerical data for large $R$ are not good enough to test this conclusively, but does not appear to directly contradict with it. 

\subsubsection{A sharp domain wall} \label{sec:d4sh}

\begin{figure}[h]
\begin{center}
\includegraphics[scale=0.8]{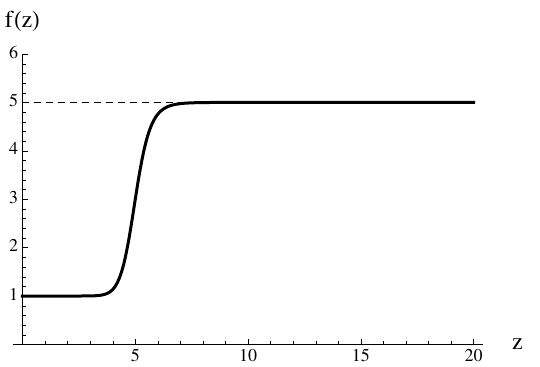} \quad
\includegraphics[scale=0.8]{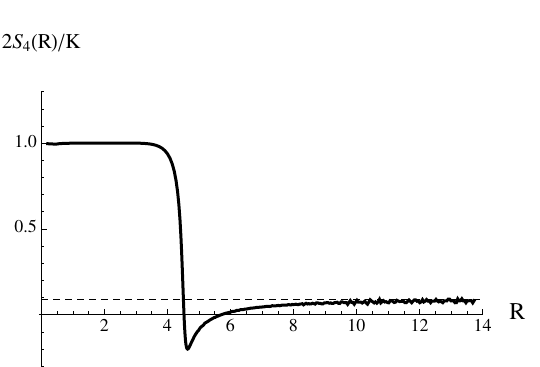} \quad
\end{center}
\caption{{\bf Left:}  A steep domain wall (toy example) in $d=4$ with $f(z) = 1 + {4 z^{15} \ov 5^{15} + z^{15}}$. {\bf Right:}  In contrast to the example of Sec.~\ref{sec:shp3d} in $d=3$, $\sS_4 (R)$ is smooth, but  is neither monotonic nor positive definite. The dash line is the expected asymptotic value for the IR fixed point. }
\label{tplot3}
\end{figure}

In the Leigh-Strassler flow discussed above, the central charges of the UV and IR fixed points are close. Our discussion in Sec.~\ref{sec:pof1} indicates that in such a situation $\sS_d (R)$ should be monotonic and positive. Now let us consider a toy example in which the domain wall is steep and the separation  of central charges is large. As indicated in Fig.~\ref{tplot3}, $\sS_4 (R)$ is neither monotonic nor positive definite.  In contrast to the $d=3$ example of Sec.~\ref{sec:shp3d}, there is no phase transition here.  

\subsubsection{Gapped phases I: GPPZ flow} \label{sec:gppz}

\begin{figure}[h]
\begin{center}
\includegraphics[scale=0.8]{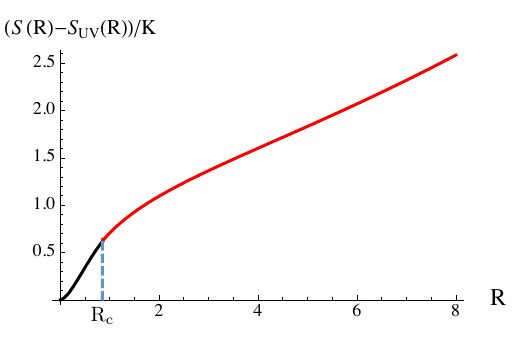} \qquad
\includegraphics[scale=0.8]{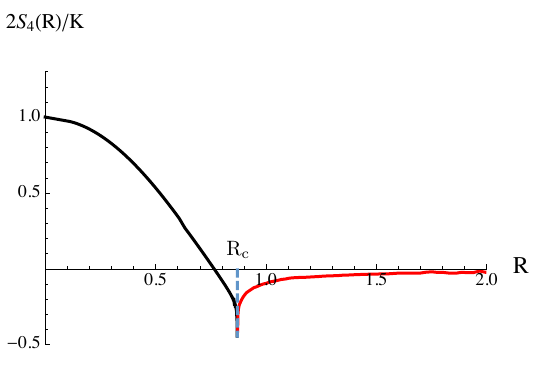}
\end{center}
\caption{ {\bf Left:} plot of entanglement entropy $S (R)- S_{\rm UV} (R)$ for GPPZ flow where $S_{\rm UV} (R)$ denotes the entanglement entropy for the UV fixed point. The short-distance divergences cancel when taking the differences. 
{\bf Right:}  plot of $\sS_4 (R)$ for GPPZ flow, which is neither positive-definite nor monotonic. The system also exhibits a ``second order phase transition'' from minimal surface of disk topology (black curve) to cylindrical topology (red curve), where $\sS_4 (R)$ remains continuous~(as far as our numerics could tell). }
\label{gppzplot}
\end{figure}

We now consider the GPPZ flow~\cite{Girardello:1999} which describes
the flow of the $\sN=4$ SYM theory to a confining theory under a mass deformation, which has 
UV dimension $\De = 3$. For this flow the metric is known analytically with 
\be  
f(z) = \le(1 + {z^2 \ov L^2} \ri)^2 \ .
\ee
Note that as $z \to \infty$, $f \sim z^4 \to \infty$. Low energy excitations of this system have a discrete spectrum with a finite mass gap (see e.g.~\cite{Bianchi:2001de}). For such a gapped phase we expect $\sS_4 (R) \to 0$ for large $R$. Fig.~\ref{gppzplot} gives the entanglement entropy and $\sS_4$ for this system.  While the finite part of the entanglement entropy appears to grow linearly with $R$ at large $R$, $\sS_4$ approaches zero from negative side. Similar to the $f(z) = 1 +z^3$ example in $d=3$ discussed in Sec.~\ref{sec:fgap}, there is a ``second order phase transition'' from minimal surface of disk topology (black curve) to cylindrical topology (red curve), where $\sS_4 (R)$ remains continuous (as far as our numerics could tell). Again as in the discussion of $f(z) = 1 +z^3$ the transition goes through the 
critical surface of  Appendix~\ref{app:A}, as indicated in Fig.~\ref{fig:cartoon2}.

\subsubsection{Gapped phases II: Coulomb branch flow} \label{sec:coul}

\begin{figure}[h]
\begin{center}
\includegraphics[scale=0.8]{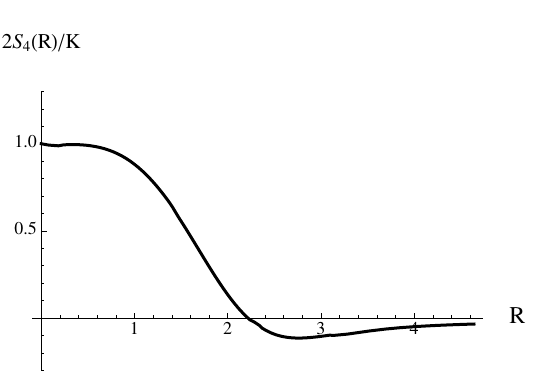} \quad
\end{center}
\caption{$\sS_4 (R)$ for the Coulomb branch flow. 
}
\label{CBplot}
\end{figure}

The Coulomb branch flow describes the spontaneous breaking of the gauge symmetry~(and also conformal symmetries) of the $\sN = 4$ SYM by a vacuum expectation value for an operator which is  bilinear in the fundamental scalars~\cite{Freedman:1999gk,Brandhuber:1999hb}. Thus here there is no operator deformation, and the end point describes a point on the Coulomb Branch.  
It involves a scalar field  of UV dimension $\De =3$. The metric is also known analytically, although through an implicit 
function
\be 
z = {1 \ov \mu} (1-v)^\ha v^{-{1 \ov 3}}, \qquad f = {(v+2)^2 \ov 9 v^{2 \ov 3}}
\ee
with $\mu$ a mass scale characterizing the expectation values of the adjoint scalars of the $\sN=4$ theory. Near $z=0$ (boundary), 
\be 
f = 1 + {2 \ov 9}  \mu^4 z^4 + \cdots, 
\ee
and near $z = \infty$, 
\be \label{fsym}
f = {4 \ov 9} \mu^2 z^2 + \cdots  \ .
\ee
While the singularity $z=\infty$ lies at a finite proper distance away, but it takes a null
geodesic an infinite time in $t$. The low energy excitations of the system include a massless Goldstone  mode from spontaneous breaking of scaling symmetry, and a continuous spectrum above a finite mass
gap (see e.g.~\cite{Bianchi:2001de}). Note that the single Goldstone mode will have a nontrivial 
$O(1)$ contribution to $\sS_4$, but is not visible in the order $O(N^2)$ we are considering. So we expect 
$\sS_4 (R) \to 0$ for large $R$ as in a gapped phase. With large $z$ behavior given by~\eqref{fsym}, there can only be minimal surface with a disk topology. The numerical result is shown in Fig.~\ref{CBplot}, with a unique minimal surface solution for all $R$. However, $\sS_4 (R)$ is neither monotonic nor positive-definite. 

\subsection{Summary}

To summarize, we find that in $d=3$ all examples (which are Lorentz invariant and satisfy the null energy condition) have a monotonic and non-negative $\sS_3 (R)$, while 
this is no longer the case for $\sS_4 (R)$ in $d=4$.  We also find that the monotonicity of $\sS_3 (R)$ is tied very closely to the null energy condition, which can be interpreted as the bulk reflection of boundary unitarity.  Thus it is tempting to conjecture it is always monotonic in the vacuum of a Lorentz-invariant, unitary QFT.

Note that although $\sS_4 (R)$ is not monotonic, given the relation~\eqref{anoi} and the recently proved $a$-theorem~\cite{Komargodski:2011vj} for $d=4$, we still always have $\sS_4 (R \to 0) > \sS_4 (R \to \infty)$.

Even if $\sS_3 (R)$ turns out to be monotonically decreasing in the vacuum we do not expect it  to remain so at a finite temperature/chemical potential. As discussed in Sec.~\ref{sec:ft} at a finite temperature we expect $\sS_d (R)$ to be proportional to the thermal entropy, i.e. proportional to the volume of the entangled region in the large $R$ limit with a positive coefficient. Similarly for a system with a Fermi surface (i.e. at a chemical potential),  we argued in Sec.~\ref{sec:fermi} that $\sS_d (R)$ should be proportional to the area of the entangled region at large $R$. While our dimensional analysis could not determine the prefactor, we expect it to be positive on physical ground. Clearly the corresponding $\sS_d (R)$ which starts as a constant for small $R$, cannot not be monotonically decreasing in either situation.  
In the holographic context, the function $f (z)$ in the spacetime metric for states of finite chemical potential  or finite temperature does satisfy~\eqref{pep2} (instead it decreases to zero at the horizon).

We also observe  first-order and second-order ``phase'' transitions
 in $\sS_d (R)$.\footnote{Note that  ``phase'' transitions in the holographic entanglement entropy have also been observed in various other contexts~\cite{Hirata:2006jx,Nishioka:2006gr,Klebanov:2007ws,bak,Pakman:2008ui,Headrick:2010zt}, but all appear to be the first-order.} By first-order, we mean $\sS_d (R)$ has a discontinuous jump, while in a second-order transition, $\sS_d (R)$ is continuous, but not smooth. When the IR is described by a conformal fixed point (corresponding to a different AdS geometry), the phase transitions appear to be first-order and  do not involve change of the topology of the minimal surface, as in a sharp domain wall in $d=3$ discussed in Sec.~\ref{sec:shp3d}. When the IR is described by a ``gapped'' phase corresponding to certain singular 
bulk IR geometries, the phase transitions (which can be first- or second-order) appear to involve change of topology of the minimal surface. Second-order transitions appear to be controlled by the critical surface  discussed in Appendix~\ref{app:A}.  Examples which exhibit a second-order transition
include the GPPZ flow~\cite{Girardello:1999} which describes the flow of the $\sN=4$ SYM theory to a confining phase under a mass deformation.  

Such phase transitions are interesting, as they  signal drastic changes of some underlying physics. 
In the example of a sharp domain wall, the system transitions very quickly (i.e. in a very short range of scales) from the UV to IR regimes, thus it appears that the entanglement entropy does not have ``time'' to respond to that quick change, which results in a discontinuity. Similarly the phase transitions in the case where the IR is a gapped phase likely signal that the system is opening a gap. Similar first-order phase transitions have been observed before in~\cite{Nishioka:2006gr,Klebanov:2007ws,bak} and were interpreted as a confinement/deconfinement transition.

\section{Conclusions and discussion}  \label{sec:conc}

In this paper we introduced a ``renormalized entanglement entropy'' $\sSS$ which appears to capture only the ``universal'' part of the entanglement entropy.  We illustrated the power of this construction by showing that  the qualitative behavior of the entanglement entropy of a system with a Fermi surface could be obtained by simple dimensional analysis. 

We also showed that (in the vacuum)  $\sSS$ has various nice features (as listed in the Introduction) which make it natural to interpret it as describing the RG flow of the entanglement entropy with distance scale. We were particularly interested in finding out whether it could be monotonic and  provide a scale-dependent measure of the number of degrees of freedom of a quantum system. 
In $d=3$ our studies of free theories and holographic systems support the conjecture that $\sS_3 (R)$ (given by~\eqref{c3}) is non-negative and monotonic for  Lorentz-invariant, unitary QFTs. 
In $d=4$, this is no longer true. We find examples in 
which $\sS_4 (R)$ (given by~\eqref{c4}) is neither monotonic nor positive-definite. 

Let us now mention some future questions:
 
\ben 

\item In our discussion $\sS_d^{(\Sig)} (R)$ was constructed by using the differential operator in~\eqref{Scen} to strip shorter-distance correlations (including cutoff dependence) away from the entanglement entropy $S^{(\Sig)} (R)$. In particular, to show its UV finiteness we had to examine the the specific structure of divergent terms in $S^{(\Sig)} (R)$. It would be interesting to have an intrinsically finite way to define it. One possibility is that one
may obtain it from certain limit of the mutual information (and their generalizations involving more than two regions).\footnote{See e.g.~\cite{Casini:2008wt,Swingle:2010jz} for discussions mutual information in general quantum field theories and of possibility of a $c$-theorem from mutual information.}

\item The monotonicity of~\eqref{c3} would imply that 
\be  \label{t1}
R {d \sS_3 (R) \ov d R} = R^2 {d^2 S \ov d R^2} < 0 \ ,
\ee
i.e. as a function $R$, $S(R)$ is a concave function.  This feels like a relation that can perhaps arise from a clever use of the strong subadditivity condition of the entanglement entropy, which is also responsible for  the monotonicity of $\sS_2 (R)$ in $d=2$~\cite{Casini:2004bw}. It is  also important to examine more examples for further confirmation or counterexamples.  
In the holographic context, for example, it would be  interesting to generalize our discussion to gravity theories with higher derivatives as in~\cite{Myers:2010xs,hung2,deboer,jimliu,sinha,Paulos,Fujita:2011we,Ogawa:2011fw} to test the robustness of the conjecture.

\item Suppose $\sS_3 (R)$ turns out to be monotonic, it is then rather curious this is no longer true for $\sS_4 (R)$ in $d=4$. 

\ben 

\item One logical possibility is that $\sS_4 (R)$ is monotonic, but the holographic examples in which it is not are pathological. The example of Sec.~\ref{sec:d4sh} is a toy model with a very sharp domain wall which is clearly artificial, while those of Sec.~\ref{sec:gppz} and~\ref{sec:coul} involve singular geometries. It would be important to completely settle this using more examples.

\item Now let us assume that $\sS_4 (R)$ is indeed not always monotonic. Then one possibility is that some modifications on the definition of $\sS_4$ are needed in order to construct a central function out of the entanglement entropy on a sphere. Another possibility is 
that the $c$-theorem for $d=4$ may require physical information which is not encoded in the 
entanglement entropy for a spherical region. One hint which points to this thinking is that  for $d=4$ or higher, the monotonicity of $\sS_d (R)$ would involve more than two derivatives on $S(R)$, e.g. for $d=4$, it amounts to 
\be \label{t3}
R^3 \p_R^3 S + R^2 \p_R^2 S  < R \p_R S   \ ,
\ee
while the strong subadditivity condition can only lead to the second derivative of $S(R)$ when applied to infinitesimally separated regions. From this perspective,  it appears likely $\sS_{d} (R)$ will also not be monotonic for $d \geq 5$. It would be good to check this explicitly. 
 It would also be interesting to understand whether non-monotonic regions signal some interesting underlying physics at those distance scales. 

\een

\item All examples of this paper have been on relativistic theories. It would be interesting to explore non-relativistic systems and also time-dependent systems.

\item  In this paper we find in holographic systems:

\ben 

\item When the difference of the central charges of the UV and IR fixed points of a QFT is sufficiently small, then $\sS_d$ is positive and monotonic for all $d$. 

\item In the small $R$ limit,  the leading $R$-dependence is given by
\be \label{asym11}
\sS_d (R) = s_d^{\rm (UV)} - A(\al) (\mu R)^{2 \al} + \cdots 
\ee
where $\al = d- \De$ for a source deformation and $\al = \De$ for a vev deformation,\footnote{This is for standard quantization. For alternative quantization one then have $\al = \De$ for source deformation and $\al = d -\De$ for vev deformation.} with $\De$ is the UV scaling dimension of the leading deformation operator (i.e. the one with the smallest $\al$). $A (\al)$ is a positive number. 

\item When the IR is described by a conformal fixed point, we expect in the large $R$ limit,  generically
\be 
\sS_d (R) = s_{d}^{\rm (IR)} + \bca {\# \ov R} + \cdots & {\rm odd \; d} \cr
                       {\# \ov R^2} + \cdots & {\rm even \; d}
                        \eca \ 
\ee
with the sign of the coefficients undetermined. 
Bur for  $\tilde \al  < \ha$ for odd $d$ and  $\tilde \al < 1$ for even $d$, where $\tilde \al = \De_{\rm IR} - d$ with $\De_{\rm IR}$ the scaling dimension of the leading irrelevant operator in the IR,  holographic systems predict that 
 \be \label{asym12}
\sS_d (R) = s_{d}^{\rm (IR)} +  {B(\tilde \al) \ov (\tilde \mu R)^{2 \tilde \al} }+ \cdots, \qquad R \to \infty
\ee
with $B (\tilde \al)$  a positive number. 

\een
It would be desirable to derive these features more generally using conformal perturbation theory.  
It is also interesting to note that the asymptotic behavior~\eqref{asym11} and~\eqref{asym12} in fact coincide with those of the interpolating $c$-function of~\cite{Komargodski:2011vj} (for $d=4$).\footnote{We thank S.~Pufu for this observation.} Likely they have a common origin.

\item In both $d=3$ and $d=4$ we observed  first-order and second-order ``phase'' transitions in $\sS_d (R)$.  For a first-order transition,  $\sS_d (R)$ has a discontinuous jump, while in a second-order transition, $\sS_d (R)$ is continuous, but not smooth. It would be interesting to understand whether  such ``phase transitions''  are due to artifacts of the the large $N$ approximation one is working with. Even if they are, their presence should still reflect some underlying features of the system at finite $N$. 
For example, as mentioned earlier, the transition in the GPPZ flow may signal the opening of a gap of the system. 

\een

\vspace{0.2in}   \centerline{\bf{Acknowledgements}} \vspace{0.2in} We thank  D.~Freedman, T.~Grover, M.~Headrick, M.~Huerta, D.~Jafferis, I.~Klebanov, R.~Myers, S.~Pufu, S.~Sachdev, T.~Senthil, B.~Swingle, T.~Takayanagi, and E.~Tonni for correspondences and conversations. Work supported in part by funds provided by the U.S. Department of Energy
(D.O.E.) under cooperative research agreement DE-FG0205ER41360.

\appendix 

\section{Induced metric and extrinsic curvature for a scalable hypersurface} \label{app:ent}

Denote the spherical coordinates in $\RR^n$ as $(r, \th_a)$ where $\th_a, a=1,\dots n-1$ denote 
 all the angular variables. Then a scalable hypersurface $\Sig$ can be specified as  
\be \label{hyper}
r = R f (\th_a)  \ ,
\ee
where $ f(\th_a)$ is a smooth function of the angular variables only and $R$ denotes the size. Clearly as we change $R$, the shape of the surface which is specified by $f$ does not change. 
Plugging~\eqref{hyper} into the metric for $\RR^n$ we find that the full flat space metric for $\RR^n$ can be written as 
\be \label{fulm}
ds^2 = (f d R + R d f)^2 + R^2 f^2 d \Om^2  = f^2 d R^2 + 2 R dR f df +  ds^2_\Sig, 
\ee 
where $d \Om^2$ is the standard metric on a unit sphere and $ds^2_\Sig$ is the induced metric 
on $\Sig$ 
\be 
ds^2_{\Sig} =  R^2 \le(f^2 d \Om^2 + (df)^2 \ri) \equiv h_{ab} d \th^a d \th^b \ .
\ee
Note that since $R^2$ appears in $h_{ab}$ as an overall prefactor, the Christoffel symbol $\Ga$ built from $h_{ab}$ is clearly independent of $R$, and thus the intrinsic covariant derivative $D_a = \p_{\th_a} + \Ga $ on $\Sig$  is independent of $R$. From~\eqref{fulm} one can also read that the standard  lapse function $N$ is independent of $R$, while the shift vector $N_a \propto R$. Thus the extrinsic curvature $K_{ab}$ which can be written as 
\be \label{defkab}
K_{ab} = \ha N^{-1} \le(\p_R h_{ab} - D_a N_b - D_b N_a \ri)  \propto R \ . 
\ee
Thus we have established~\eqref{scalo}. Since in $F$ of~\eqref{locE} all indices have to contracted and there are an even number of $K_{ab}$, thus $F$ can only depend on $R$ through $R^{-2n}$ 
for $n$ some non-negative integer. This then establishes~\eqref{divv}.\footnote{For theories breaking parity it is possible to have terms with different powers from those indicated in~\eqref{divv}~\cite{tarun}, but they have negative powers of $R$. From our discussion in the main text, they will not give rise to divergent terms.} Note that $n=2$ is a bit special as in this case, both the curvature $K$ and $h$ have only a single component, 
but the same conclusion applies. 

For illustration, let us also give an example of a surface which is not scalable. Consider in $\RR^2$
a curve specified by 
\be 
x^2 + {y^4 \ov b^2} = R^2 \ 
\ee
whose shape clearly changes with $R$. For this curve there is an additional dimensional parameter $b$, and our previous discussion does not apply.


\section{Details of the numerical calculation of $\sS_3 (R)$ for a free massive scalar} \label{app:num}

We start with the Hamiltonian for a free massive scalar in $d=3$
\begin{equation}
H=\frac{1}{2}\int d^2 x \ \le[\Pi^2+ (\nabla {\phi})^2 +m^2 \phi^2 \ri]\,.
\end{equation}
where $\Pi$ is canonical momentum for $\phi$. 
In terms of the Fourier transform of $\phi$ and $\Pi$ in the angular $\th$ direction
\begin{eqnarray}
\phi_0 &=& \sqrt{\frac{r}{2\pi}}\int d\theta \;  \phi(\theta,r), \qquad  \phi_l  + i \phi_{-l} =\sqrt{\frac{r}{\pi}}\int d\theta ~e^{i l\theta} \phi(\theta,r), \quad l > 0\\
\Pi_0 &=&\sqrt{\frac{r}{2\pi}}\int d\theta \;  \Pi(\theta,r) , \qquad \Pi_l + i \Pi_{-l} = \sqrt{\frac{r}{\pi}}\int d\theta~ e^{i l\theta} \Pi(\theta,r) , \quad l > 0
\end{eqnarray}
the Hamiltonian then can be written as $H=\sum_{l=-\infty}^\infty H_l$ with 
\begin{equation}
 H_l=\frac{1}{2}\int_0^{\infty} dr \ \le[\Pi_l^2+r\partial_r\le(\frac{\phi_l}{\sqrt{r}}\ri)^2+m^2 \phi_l^2+\frac{l^2}{r^2} \phi_l^2 \ri]\,.
\label{hscalar}
\end{equation}
Also note that $\phi_l, \Pi_l$ satisfy the canonical commutation relation 
\begin{equation}
 [\phi_l(r),\Pi_{ l'}(r^{\prime})]=i \delta_{ll^{\prime}}\delta(r-r^{\prime} )\,.
\end{equation}
We discretize~\eqref{hscalar} with a uniform lattice in the radial direction:
\begin{equation}
 H_l=\frac{1}{2a} \sum_{j=1}^N \ \le[\Pi_l(j)^2+\le(j+\ha\ri)\le[\frac{\phi_l(j)}{\sqrt{j}}-\frac{\phi_l(j+1)}{\sqrt{j+1}}\ri]^2+m^2 \phi_l^2(j)+\frac{l^2}{j^2} \phi_l(j)^2 \ri]\,,
\label{hscalardisc}
\end{equation}
where $a$ is the lattice spacing and $r=j a$ and we introduced an IR cutoff $\Lam_{IR}=Na$. The radius of the disk is taken to be:
\be
R=\le(n+\ha\ri) a \ . \label{rchoice}
\ee

The Hamiltonians~\eqref{hscalardisc} can be written in a general form
\be
H=\frac{1}{2}\sum_{i=1}^N \Pi _{i}^{2}+\frac{1}{2}\sum_{ij=1}^N \phi _{i}K_{ij}\phi_{j} \  ,
\ee
and we are interested in the entanglement entropy $S(n,N)$ for the subset of degrees of freedom  $\phi_{\al}, \al =1,2, \cdots n < N$ when the system is in the vacuum. The problem was solved in
~\cite{Srednicki:1993im} and the result can be written as follows. 
Decompose $\Omega\equiv\sqrt{K}$ as 
\be
\Omega=\begin{pmatrix}A&B\\B^T&C\end{pmatrix}\,
\ee
where $A$ has rank $n$ and let 
\be \label{brpt}
\beta=\frac{1}{2} B^TA^{-1} B\,,\qquad
\beta^\prime=\frac{1}{\sqrt{C-\beta}}\beta \frac{1}{\sqrt{C-\beta}}\,,\qquad
\Xi=\frac{\beta^\prime}{1+\sqrt{1-\beta^{\prime 2}}} \,.
\ee
Then, $S(n,N)$ can be written in terms of the $(N-n)\times(N-n)$ matrix $\Xi$ as 
\be \label{welc}
S (n,N)=-\tr \left [\log (1-\Xi ) +\frac{\Xi}{1-\Xi} \log \Xi \right ]\,.
\ee

For our case:
\bea
K_l^{11}&=&\frac{3}{2}+j^2+m^2\,,\cr
K_l^{jj}&=&2+\frac{l^2}{j^2}+m^2\,,\cr
K_l^{j,j+1}&=&-\frac{j+1/2}{\sqrt{j(j+1)}}=K_n^{j+1,j}
\label{defk}
\eea
and the entanglement entropy for a spherical region of radius $R$ can then be written as 
\be
S(R) = \sum_{l=-\infty}^{\infty}  S_l (n, N) \, 
\ee
where $S_l (n,N)$ is given by~\eqref{welc} with $\Xi_l$ obtained from~\eqref{defk} via~\eqref{brpt} for each $l$.   

To get $S_l (n,N)$ we need the eigenvalues of $\Xi_l$. It can be checked that all square roots and inversions are well defined, and all the eigenvalues of $\Xi_l$ lie between $0$ and $1$. One can also show that the sum over $l$ converges at fixed $n,\, N$.

For $m=0$, $S(R)$ entropy should have the form
\be
S(R)=c_1 {R \ov a} - s_3 + O\le({a \ov R}, {R\ov \Lam_{IR}}\ri)
\ee
where $s_3$ is known analytically~\cite{dowker} 
\be
s_3={1\ov 16}\le(2\log2- {3\zeta(3)\ov \pi^2}\ri)\approx 0.0638 \ .
\ee
From our numerical calculations we get for $m=0$
\be
c_1=0.4643821 \qquad s_3=0.0635\pm0.0004
\ee
The calculations were performed with the choice of IR cutoff $N=200$. To avoid the boundary effects from both the UV (small $n$) and IR (large $n$) cutoffs, we needed to restrict to  the range $10\leq n \leq 45$.  We require $10^{-6}$ absolute accuracy for the result of the entropy calculation in order to be able to extract $\sS_3(R)$ with satisfactory precision. To achieve this we followed the analysis of \cite{Lohmayer:2009sq} which shows that the finite volume corrections (accounting for a finite $N$) go as $N^{-(2l+2)}$. With our choice $N=200$, they are thus negligible for $l\geq3$. For $l<3$ we determine the coefficient of the correction by doing the calculation for several choices of $N>200$ and extrapolate to infinite $N$. We refer the reader to~\cite{Lohmayer:2009sq} for details.

At finite $m$, 
 given the limited range of $10\leq n \leq 45$ (thus limited range of $R$) for which we can do calculations for a given $m$, we extend the range of $mR$ by 
  working with different choices of $m$ in lattice units. Since $\sS_3 (R)$ can only depend on 
  $mR$ in the continuum limit, data points from different choices of $m$ should collapse into a single curve, and they indeed do. The results are shown in Fig.~\ref{fig:frm} of Sec.~\ref{sec:free}. Choosing $1/m=20 a$ enabled us to cover the range $0.55<mR<2.25$, while $1/m=40a,\ 120a$ cover the range $0.275<mR<1.125$ and $0.09<mR<0.375$ respectively. 
  In Fig.~\ref{fig:frm} one can see that in the overlapping regimes the data points agree, justifying the continuum extrapolation of the lattice results, i.e. the $\sS_3(mR)$ determined here is independent of the discretization. 

\section{Cylinder-like solutions} \label{app:A}

In this Appendix we discuss when a cylinder-like solution minimal surface could appear. We will show that for a bounded $f$, the minimal surface solution is always disk-like, while when $f(z) \sim z^n$ for large $z$, a cylinder-like solution can appear only when $n > 2$. 

Extremizing the minimal surface action~\eqref{minar} leads to the 
 equation of motion 
 \be \label{eom1}
(d-2)  {1 \ov f }  + (d-1) {\rho \rho'  \ov z} 
= {\rho } \sqrt{\rho'^2 + {1 \ov f}} \p_z \le( {\rho' \ov \sqrt{\rho'^2 + {1 \ov f}}} \ri) \ .
\ee

Let us first consider a bounded $f$, i.e. with asymptotic behavior 
\be 
f (z) = f_0 + f_1 z^{-m} + \cdots , \quad m > 0, \quad z \to \infty \ .
\ee 
Then with a cylinder-like solution 
\be \label{cym}
\rho (z) = \rho_0 + \rho_1 z^{\al} + \cdots ,  \quad  \al < 0, \quad z \to \infty \ , 
\ee
in the equation of motion~\eqref{eom1}, the leading term on the left hand side (LHS) is of order $O(1)$ while that on the right hand (RHS) is of order $O(z^{\al -2})$, and  thus a solution is not possible. 

Now let us consider an unbounded $f(z)$ with the large $z$ behavior
\be 
f^{-1} =  a z^{-n} \le(1 + \cdots \ri), \qquad  n >0, \quad z \to \infty \ .
\ee
With~\eqref{cym}, in the equation of motion~\eqref{eom1}, the leading term on the left hand side is 
\be \label{lhs}
{\rm LHS} = (d-2) a z^{-n}   + (d-1) \rho_0 \rho_1 \al z^{\al -2}  + \cdots \ . 
\ee
The leading behavior on the RHS is given by 
\be \label{rhs}
{\rm RHS} = \bca  -\ha \rho_0 \rho_1 \al  (2-n-2 \al) z^{-n - \al} & \al > 1-{n \ov 2} \cr
                                \ha \rho_0 \rho_1 \al  (n+2 \al-2) z^{ \al-2} & \al < 1-{n \ov 2} \cr
                                 \sim z^{-\beta}, \quad \beta > {n \ov 2} + 1   & \al = 1-{n \ov 2} 
     \eca
\ee 
Comparing~\eqref{lhs} and~\eqref{rhs} we find that the leading order equation can only be satisfied 
for the case of the second line of~\eqref{rhs} with
\be \label{sol1}
\al = 2 -n \ .
\ee
Note that since $\al < 0$, this requires $n > 2$. We also find that the equation of motion requires 
\be 
\rho_1 =  {2 (d-2) a \ov \rho_0 (n-2) (n+2d-4)}   \ .
\ee
We thus conclude that for $n > 2$, a cylinder-like solution could exist with the large $z$ behavior given by
\be \label{cym1}
\rho (z) = \rho_0 + {2 (d-2) a \ov \rho_0 (n-2) (n+2d-4)}   z^{2-n} + \cdots , \quad z \to \infty \ .
\ee

We notice that in~\eqref{cym1} the coefficient before the second term becomes singular when $\rho_0 \to 0$. This implies if there is a solution with $\rho_0 = 0$ it should comes with a larger power than~\eqref{sol1}. 
Writing for large $z$, $\rho(z) = \rho_1' z^{\al'} + \cdots$ with $\al' < 0$, 
we again find that a solution is possible only for $n >2$, for which the leading behavior of the solution is given by
\be \label{cms}
\rho (z) = \sqrt{2 (d-2) a \ov (d-1) (n-2)}  z^{1-{n \ov 2}} \le(1 + \cdots \ri) \ .
\ee
Equation~\eqref{cms} describes a solution in which the minimal surface just closes off at $z = \infty$. 
Note that~\eqref{cms} can be considered as the $\rho_0 \to 0$ limit of~\eqref{cym1} in the following sense. Equation~\eqref{cym1} was derived assuming that the second term is much smaller than the first term, i.e. for 
\be 
z^{n-2} \gg {1 \ov \rho_0^2} {2 (d-2) a \ov (n-2) (n+2d-4)} \equiv z_0^{n-2}  \ .
\ee 
When $\rho_0^2 \ll a$, equation~\eqref{cms} is valid in the region $1 \ll z < z_0$, while one has~\eqref{cym1} for $z \gg z_0$. In the limit $\rho_0 \to 0$, $z_0 \to \infty$, and thus~\eqref{cms} becomes valid in the full large $z$ region. 

The solution~\eqref{cms} can be considered as the critical surface lying at the boundary between the spaces  
of disk-like and cylinder-like solutions. As discussed in Sec.~\ref{sec:num}, indeed it
appears to control the transition point of certain ``second order phase transitions'' from minimal surface of disk topology to cylindrical topology.

\end{document}